\documentclass[aps, prd, amsmath, floats, floatfix, superscriptaddress,twocolumn,
nofootinbib,eqsecnum]{revtex4}
\pdfoutput=1
\usepackage{graphicx}
\usepackage{amsmath,amssymb}
\usepackage{hyperref}
\usepackage{amsfonts}
\usepackage{xspace} 
\usepackage{color}
\usepackage{dcolumn}
\usepackage{bm}

\def\laq{\raise 0.4ex\hbox{$<$}\kern -0.8em\lower 0.62ex\hbox{$\sim$}}
\def\gaq{\raise 0.4ex\hbox{$>$}\kern -0.7em\lower 0.62ex\hbox{$\sim$}}

\newlength{\sizeonefig}
\newlength{\sizetwofig}
\newlength{\sizeonefigb}
\newlength{\sizetwofigb}
\newcommand*{\mysim}{\mathord{\sim}}
\newcommand*{\li}{\textsc{LALInference}}
\newcommand*{\mcmc}{\textsc{lalinference\_mcmc}}
\newcommand*{\gstlal}{\textsc{gstlal\_inspiral}}
\newcommand*{\mc}{\mathord{\cal M}}
\newcommand*{\q}{\mathord{q}}
\newcommand*{\thetaa}{\mathord{\ensuremath{\theta_{\text{NS}_1}}}}
\newcommand*{\thetab}{\mathord{\ensuremath{\theta_{\text{NS}_2}}}}
\newcommand*{\phia}{\mathord{\ensuremath{\phi_1}}}
\newcommand*{\phib}{\mathord{\ensuremath{\phi_2}}}
\newcommand*{\tilta}{\mathord{\ensuremath{\theta_{\text{LS}_1}}}}
\newcommand*{\tiltb}{\mathord{\ensuremath{\theta_{\text{LS}_2}}}}
\newcommand*{\thetajn}{\mathord{\ensuremath{\theta_{\text{JN}}}}}
\newcommand*{\thetajl}{\mathord{\ensuremath{\theta_{\text{JL}}}}}
\newcommand*{\phiab}{\mathord{\ensuremath{\phi_{12}}}}
\newcommand*{\phijl}{\mathord{\ensuremath{\phi_\text{JL}}}}
\newcommand*{\psil}{\mathord{\ensuremath{\psi_\text{L}}}}
\newcommand*{\psij}{\mathord{\ensuremath{\psi_\text{J}}}}

\newcommand*{\bnsfactor}{7.6}
\newcommand*{\nsbhfactor}{11}
\newcommand*{\bbhfactor}{3.7}
\newcommand*{\factorrange}{3.7{\rm -}11}

\allowdisplaybreaks

\begin{document}

\title{A more effective coordinate system for parameter estimation of precessing compact binaries from gravitational waves}

\author{Benjamin Farr}
\email{bfarr@u.northwestern.edu}
\affiliation{Center for Interdisciplinary Exploration and Research in
    Astrophysics (CIERA) \& Dept. of Physics and Astronomy, Northwestern
    University, 2145 Sheridan Rd, Evanston, IL 60208, USA.}
\affiliation{School of Physics and Astronomy, University of Birmingham,
    Edgbaston, Birmingham B15 2TT, UK}

\author{Evan Ochsner}
\email{evano@gravity.phys.uwm.edu}
\affiliation{Center for Gravitation and Cosmology, University of
    Wisconsin-Milwaukee, Milwaukee, WI 53201, USA }

\author{Will M. Farr}
\email{w.farr@bham.ac.uk}
\affiliation{School of Physics and Astronomy, University of Birmingham,
    Edgbaston, Birmingham B15 2TT, UK}

\author{Richard O'Shaughnessy}
\affiliation{Center for Gravitation and Cosmology, University of
    Wisconsin-Milwaukee, Milwaukee, WI 53201, USA }

\begin{abstract}
  Ground-based gravitational wave detectors are sensitive to a narrow range of
  frequencies, effectively taking a snapshot of merging compact-object binary
  dynamics just before merger.  We demonstrate that by adopting analysis
  parameters that naturally characterize this `picture', the physical
  parameters of the system can be extracted more efficiently from the
  gravitational wave data, and interpreted more easily.  We assess the
  performance of MCMC parameter estimation in this physically intuitive
  coordinate system, defined by (a) a frame anchored on the binary's spins and
  orbital angular momentum and (b) a time at which the detectors are most
  sensitive to the binary's gravitational wave emission.  Using anticipated
  noise curves for the advanced-generation LIGO and Virgo gravitational wave
  detectors, we find that this careful choice of reference frame and reference
  time significantly improves parameter estimation efficiency for BNS, NS-BH,
  and BBH signals.
\end{abstract}

\date{\today}

\maketitle

\section{Introduction}
\label{sec:intro}

The Advanced LIGO (aLIGO) gravitational-wave detectors are expected to come
online in 2015~\cite{Harry:2010zz}, with Advanced Virgo (AdV) following in
2016~\cite{aVirgo}.  They are expected to directly detect gravitational waves
and, once they reach design sensitivity, could detect tens of events per
year~\cite{Abadie:2010cf} from the coalescence of compact binaries composed of
neutron stars (NSs) and/or black holes (BHs). Black holes in particular are
expected to have large spin~\cite{McClintock:2011zq}, so properly modeling
spins will be especially important for NS-BH and BBH binaries.  Spin-induced
precession, caused when one or both spin vectors are misaligned with the
orbital angular momentum, is a particularly challenging effect to model, as it
induces amplitude and phase modulations, changes the relative strength of the
two waveform polarizations, and complicates a spin-weighted spherical harmonic
mode decomposition of the waveform~\cite{Kidder:1995zr,Arun:2008kb}.

In an important early work, Apostolatos~et.~al.~\cite{Apostolatos:1994mx} laid
out a ``simple precession'' model for the evolution and gravitational wave
emission of compact binaries with generic spins.  With the exception of rare
cases of transitional precession, or the simpler special case of spins aligned
or anti-aligned with the orbital angular momentum, the orbital angular momentum
vector of the binary will precess on a cone about the total angular momentum.
Since this and other important early work~\cite{Kidder:1992fr,Kidder:1995zr},
there have been a number of improvements to waveform models from spinning
systems.  This has included deriving and studying further post-Newtonian (PN)
corrections to the waveform dynamics and
phasing~\cite{Faye:2006gx,Blanchet:2006gy,Marsat:2012fn,Bohe:2013cla},
amplitude~\cite{Kidder:1995zr,Arun:2008kb} and precession
equations~\cite{Bohe:2012mr}, the development of spinning, precessing
inspiral-merger-ringdown
waveforms~\cite{Sturani:2010ju,Schmidt:2012rh,Hannam:2013oca,Taracchini:2013rva}
and a frequency-domain precessing waveform model~\cite{Lundgren:2013jla}, and
efforts to track precessional motion and disentangle it from other dynamical
effects~\cite{Schmidt:2010it,Boyle:2011gg,O'Shaughnessy:2011fx,Ochsner:2012dj,Boyle:2013nka}.
Despite all of these refinements, precessing waveforms still qualitatively
match the Apostolatos~et.~al. picture of the orbital angular momentum moving
along a precession cone that slowly grows due to radiation reaction.

Although suboptimal, matched filter searches with non-spinning or aligned-spin
templates could still detect GW signals from spinning, precessing binaries,
albeit with significant parameter bias~\cite{Harry:2013tca}.  If a detection
can be made, then the \li~module of LAL~\cite{LAL} can be used for a focused
parameter estimation followup effort~\cite{Aasi:2013jjl}.  \li~is a suite of
routines to perform Bayesian inference techniques such as Markov chain Monte
Carlo (MCMC) and nested sampling on gravitational wave detector data.  Bayesian
inference has proven adept at sampling the 11--15 dimensional parameter space
of circularized compact binary mergers with non-negligible
spin~\cite{2008ApJ...688L..61V,2009CQGra..26k4007R,2013arXiv1308.4704O,Aasi:2013jjl,2014arXiv1403.0544O,2014arXiv1403.0129V}.

Nonetheless, there is a significant computational cost to performing Bayesian
inference.  Typically, waveform generation is the dominant cost.  Performing
parameter estimation with MCMC on real or simulated data typically requires
generating several million CBC waveforms to produce $\simeq 1000$ independent
samples from the posterior probability density on parameter space.  Depending
on the computational cost of generating the waveforms, this can take hours to
weeks or longer to complete.  The latency of such analyses must, at the very
least, be low enough to keep up with potential advanced-detector trigger rates.
If these analyses are completed within hours, then they can potentially play a
critical role in the search for electromagnetic counterparts to BNS or NS-BH GW
signals.

The latency of MCMC analyses, assuming fixed waveform generation costs, are
governed by the sampling efficiency of the chains, which are intimately tied to
the parameterization and proposal distributions that are used.  In an extreme
limit, local one-dimensional jumps and distorted coordinates require the
Markov chain to jump slowly through tightly-correlated, twisting paths
in parameter space.

In this work we demonstrate that MCMC-based parameter estimation is
significantly more efficient when we adopt coordinates well-adapted to the
dynamics of the precessing binary.  We parameterize the spin and orbital
angular momentum degrees of freedom of a precessing binary with a set of angles
describing the simple precession cone model of
Apostolatos~et.~al.~\cite{Apostolatos:1994mx}.  We use the inclination of the
\emph{total} angular momentum to the line of sight, the azimuthal position of
orbital angular momentum $L_N$ on this cone at some reference point, three
angles describing the orientation of the spins relative to $L_N$ and each other
at the reference point, and the magnitude of each spin.  Critically, we also
choose the GW frequency (i.e. twice the instantaneous orbital frequency)
$f_{\rm ref}$ at which these angles are defined.  Choosing such a reference
frequency near the peak sensitivity of the detector also dramatically improves
the convergence of Bayesian parameter estimation methods.  We find that the new
parameterization and a suitable choice of $f_{\rm ref}$ can decrease the
``mixing'' or autocorrelation time in the stochastic parameter sampling by a
factor of $\factorrange{}$ for the cases considered here.  We find the shortest
autocorrelation times and best parameter constraints occur for $f_{\rm ref}
\simeq$ 70--100 Hz.

This work is organized as follows.  In Section \ref{sec:coords}, we briefly
review the dynamics of precessing binaries and the commonly used ``radiation
frame'' coordinates, and we introduce our well-adapted coordinate system for
generic precessing binaries.  In Section \ref{sec:MCMC} we very briefly
describe the parallel-tempered Markov-Chain Monte Carlo parameter estimation
strategy adopted in this work, emphasizing why well-adapted coordinates improve
its performance.  In Section \ref{sec:results}, we compare the results of
parameter estimation calculations performed using radiation frame coordinates
and our well-adapted coordinates.  We show that MCMC calculations using the
well-adapted coordinates converge more efficiently, providing reliable results
in less time.  Furthermore, our well-adapted coordinates correspond to
physically pertinent, well-constrained observables, so the posteriors for
astrophysically interesting angles such as spin tilts are obtained directly
without further (expensive) post-processing.  We finish with some concluding
remarks in Section~\ref{sec:conclusions}.

\section{Coordinates for precessing binaries}
\label{sec:coords}

\subsection{Evolution equations for precessing PN binaries}

Here for convenience we briefly review the formalism for generating PN
precessing waveforms.  For more details, we refer the reader
to~\cite{Buonanno:2009zt} (and references therein) for a summary of various
non-spinning PN waveforms. For a generalization to spinning waveforms, the
reader may refer to~\cite{Kidder:1995zr,Pan:2003qt,Buonanno:2004yd,Arun:2008kb}
among others.

The gravitational wave strain observed by a detector is given by\footnote{
	We note that this expression assumes that the sky position of the source 
	remains constant, i.e. that the Earth rotates by a negligible amount over 
	the duration of the signal. As NS-BH GW signals evolve from 10 Hz to 
	coalescence in $\lesssim 5$ minutes, this is true to a very good approximation.}
\begin{equation} \label{eq:measured_strain}
    h(t) = F_+ \ h_+(t) + F_\times \ h_\times(t)\ ,
\end{equation}
where $F_+$ and $F_\times$ are antenna pattern functions describing the
detector and $h_+(t)$ and $h_\times(t)$ are the gravitational wave
polarizations. For a compact binary evolving along a series of quasi-circular
orbits, these functions have the following form at leading order
\begin{eqnarray} 
    h_+(t) &=& - \frac{2\,M\,\eta}{D_L} v(t)^2 \left(1 + \left( \hat{L} \cdot \hat{N} \right)^2 \right)
    \, \cos 2 \phi(t)\ ,\label{eq:plus_pol}\\
    h_+(t) &=& - \frac{2\,M\,\eta}{D_L}\,v(t)^2\, 2 \left( \hat{L} \cdot \hat{N} \right)\, \sin 2 \phi(t)\ .\label{eq:cross_pol}
\end{eqnarray}
Here $D_L$ is the luminosity distance to the binary, $M$ is the total mass, 
$\eta = m_1\,m_2 / M^2$ is the symmetric mass ratio, 
$\hat{L}$ is the direction of orbital angular momentum,
$\hat{N}$ is the direction of GW propagation,
$\phi$ is the orbital phase of the binary and 
$v = (2 \pi M F_{\rm orb})^{1/3}$ is the ``characteristic velocity'' PN
expansion parameter (with $F_{\rm orb}$ the orbital frequency).
Higher order corrections to the polarizations valid for precessing binaries can
be found in~\cite{Kidder:1995zr,Arun:2008kb}.  For our purposes, the important
point is that these polarizations depend only on the masses, the inclination
between the orbital angular momentum and the line of sight, and the
time-dependent phasing and frequency parameters $\phi(t)$ and $v(t)$.\footnote{
    At higher order, the spins also enter the polarizations. Our main point is
    that Eqs.~(\ref{eq:energy_balance})--(\ref{eq:dS2dt}) are sufficient to
    compute the polarizations, and this is true at all PN orders.}

For compact binaries on quasi-circular orbits, $\phi(t)$ and $v(t)$ can be
computed via the energy balance equation.  PN expansions are known for both the
binding energy of the binary, $E$, and the gravitational wave luminosity
(commonly called the ``flux''), ${\cal F}$.  We assume that emission of
gravitational waves accounts for all of the loss of binding energy and a simple
use of the chain rule provides a differential equation for the evolution of
$v(t)$:
\begin{equation} \label{eq:energy_balance}
    -\frac{dE}{dt} = \frac{dE}{dv}\,\frac{dv}{dt} = {\cal F} \quad \Longrightarrow \quad \frac{dv}{dt} = - \frac{\cal F}{dE/dv}\ .
\end{equation}
Since the derivative of the orbital phase is simply the (angular) orbital
frequency, we trivially obtain a coupled differential equation for the phase
$\phi(t)$:
\begin{equation} \label{eq:dphidt}
    \frac{d\phi}{dt} = 2\,\pi\,F_{\rm orb} = \frac{v^3}{M}\ .
\end{equation}

For non-spinning binaries, these are the only equations needed to evolve the
orbital dynamics and compute a gravitational waveform.
Eqs.~(\ref{eq:energy_balance}) and (\ref{eq:dphidt}) are integrated to compute
$v(t)$ and $\phi(t)$, which are then plugged back into Eqs.~(\ref{eq:plus_pol})
and (\ref{eq:cross_pol}) to obtain the waveform.  There are a number of
different ways to solve these differential equations which are equivalent up to
the PN order of $E$ and ${\cal F}$, but differ in truncation error at the next,
unknown PN order. These different methods for solving the energy balance
equation are known as PN \emph{approximants}.  For example, $E$ and ${\cal F}$
are known as Taylor series in $v$, so the right hand side of
Eq.~\ref{eq:energy_balance} is a rational function of $v$.  One could keep it
in this form (referred to as the TaylorT1 approximant), or re-expand it as a
Taylor series (TaylorT4). In this work we will use TaylorT4, but our results
are equally applicable to any PN approximant.  See~\cite{Buonanno:2009zt} for a
summary of the various PN approximants,
and~\cite{Lundgren:2013jla,Nitz:2013mxa} for two newly-proposed approximants.

For precessing binaries, we note that there are spin corrections proportional
to $\hat{S}_{1,2} \cdot \hat{L}$ and $\hat{S}_1 \cdot \hat{S}_2$ in $E$ and
${\cal F}$. Additionally, we have already noted that $\hat{L} \cdot \hat{N}$
appears in the polarizations. The orientations of these vectors can change over
time for precessing binaries, and so they must be computed as functions to
time, along with $v(t)$ and $\phi(t)$.  Their evolution is given by the
precession equations~\cite{Apostolatos:1994mx,Kidder:1995zr}.
\begin{widetext}
\begin{eqnarray}
\frac{d\hat{\mathbf{L}}}{dt} &=& \frac{v^6}{2 M^3} \left\{ \left[ \left( 4 + 3 \frac{m_2}{m_1} \right) \mathbf{S}_1 
	+  \left( 4 + 3 \frac{m_1}{m_2} \right) \mathbf{S}_2 \right]
	- \frac{3 v}{M^2 \eta} \left[ \left( \mathbf{S}_2 \cdot \hat{\mathbf{L}} \right) \mathbf{S}_1 
	+ \left( \mathbf{S}_1 \cdot \hat{\mathbf{L}} \right) \mathbf{S}_2 \right] \right\} \times \hat{\mathbf{L}}\ ,\label{eq:dLdt}\\
\frac{d\mathbf{S}_1}{dt} &=& \frac{v^5}{2 M} \left\{ \left( 4 + 3 \frac{m_2}{m_1} \right) \hat{\mathbf{L}} 
	+ \frac{v}{M^2} \left[ \mathbf{S}_2 - 3 \left( \mathbf{S}_2 \cdot \hat{\mathbf{L}} \right) \hat{\mathbf{L}}\right] \right\}
	\times \mathbf{S}_1 \ ,\label{eq:dS1dt}\\
\frac{d\mathbf{S}_2}{dt} &=& \frac{v^5}{2 M} \left\{ \left( 4 + 3 \frac{m_1}{m_2} \right) \hat{\mathbf{L}} 
	+ \frac{v}{M^2} \left[ \mathbf{S}_1 - 3 \left( \mathbf{S}_1 \cdot \hat{\mathbf{L}} \right) \hat{\mathbf{L}}\right] \right\}
	\times \mathbf{S}_2\ .\label{eq:dS2dt}
\end{eqnarray}
\end{widetext}
To generate a gravitational wave signal from a precessing binary, we first
solve for its orbit and spin [Eqs.~(\ref{eq:energy_balance})--(\ref{eq:dS2dt})],
then substitute these kinematic quantities into Eqs.~(\ref{eq:plus_pol}) and
(\ref{eq:cross_pol}).  

Higher order polarization expressions valid for precessing binaries can be
found in~\cite{Arun:2008kb}.  While in this work we report results using only
leading-order polarizations, related work by
\cite{O'Shaughnessy:2013vma,O'Shaughnessy:2014dka} suggests our conclusions
will also hold for higher-order polarizations.

\subsection{Behavior of precessing waveforms}
\label{subsec:precessing_behavior}

The behavior of precessing binaries was first laid out in significant
qualitative and quantitative detail in~\cite{Apostolatos:1994mx}.  To
summarize, they find that almost all spinning binary configurations would
undergo simple precession. This means that the direction of total angular
momentum, $\hat{J}$, remains very nearly fixed and the orbital angular momentum
vector moves on a cone about $\hat{J}$.  The angle between $\mathbf{L}$ and the
total spin, $\mathbf{S} = \mathbf{S}_1 + \mathbf{S}_2$, will remain nearly
constant throughout the binary evolution.  The magnitude of $\mathbf{L}$ will
slowly decrease due to radiation reaction, and since its angle with
$\mathbf{S}$ does not change appreciably, this means the opening angle of the
$\mathbf{L}$ precession cone will slowly grow as the binary inspirals.

There are two types of special cases of spinning binary configurations which do
not obey simple precession.  First, if both spins are aligned and/or
anti-aligned with the orbital angular momentum, then the binary will not
precess and the orbital angular momentum will remain in a fixed direction.  It
is clear from Eqs.~(\ref{eq:dLdt})--(\ref{eq:dS2dt}) that the direction of all
of these vectors will be constant when they are all parallel.  Second,
transitional precession can occur if the binary has large spins which are
nearly, but not perfectly anti-aligned with the orbital angular momentum.  In
this case, the binary will initially be in a state where the orbital angular
momentum is the dominant contribution to $\mathbf{J}$.\footnote{Since
    $\mathbf{L} \simeq \mathbf{r} \times \mathbf{p}$, the orbital angular
    momentum will always dominate the spin angular momentum for sufficiently
wide binary separations $r$.} Gravitational wave emission will decrease the
magnitude of $\mathbf{L}$, but will not change the magnitude of $\mathbf{S}_1$
and $\mathbf{S}_2$. If it decreases such that $|\mathbf{L}| \lesssim
|\mathbf{S}|$, then the direction of $\hat{J}$ will change from being near
$\hat{L}$ to near $\hat{S}$. This brief period when $|\mathbf{J}| \approx 0$
and the direction of $\hat{J}$ changes rapidly is known as transitional
precession. Note that it is preceded and proceeded by periods of simple
precession.

\begin{figure*}
\includegraphics[width=0.45\textwidth]{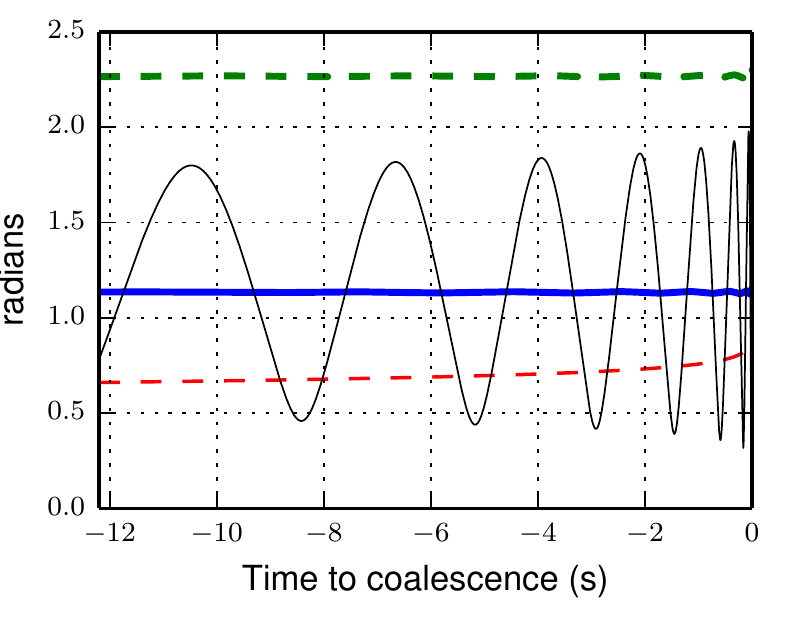}
\includegraphics[width=0.45\textwidth]{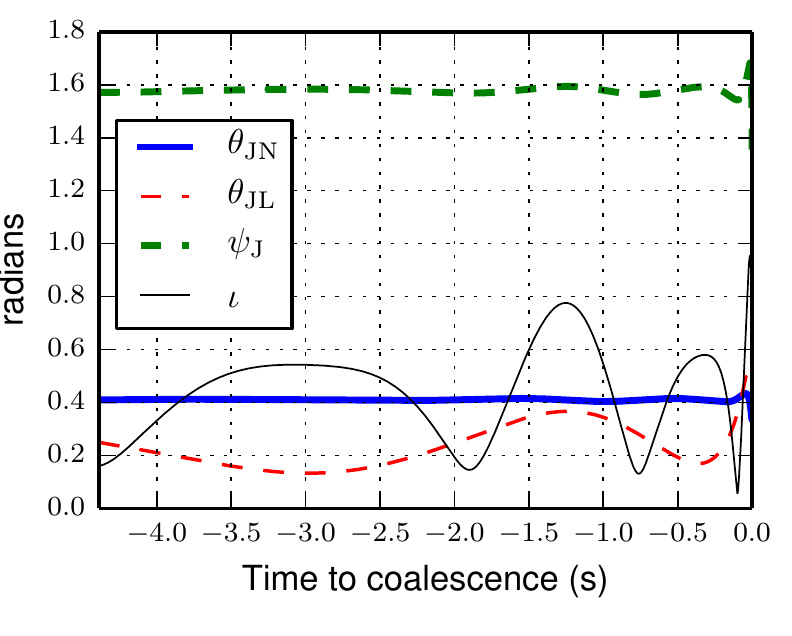}
\includegraphics[width=0.45\textwidth]{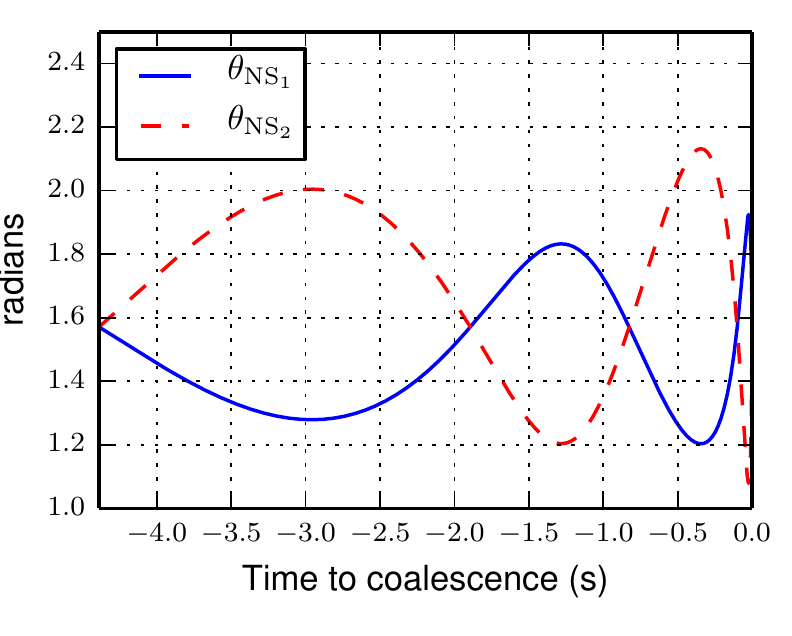}
\includegraphics[width=0.45\textwidth]{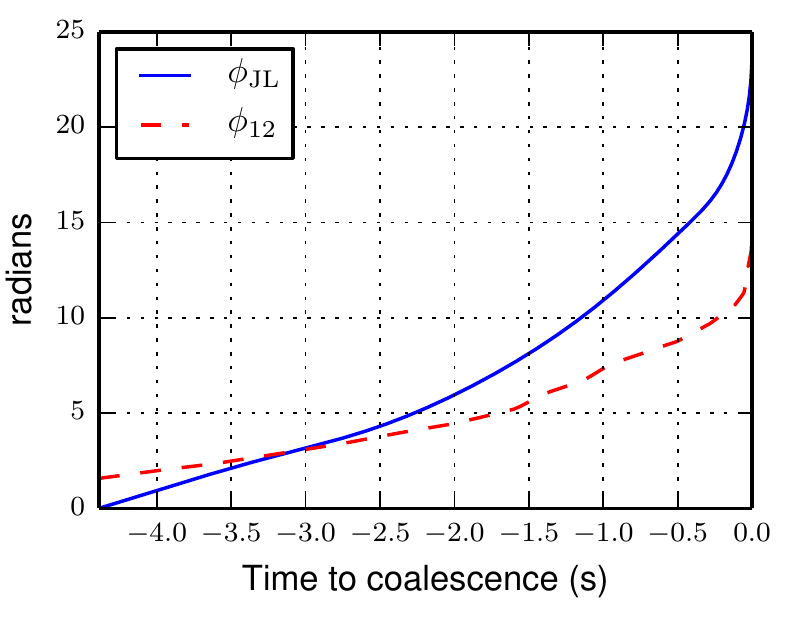}
\caption{\textbf{Fiducial binary orientation}: We plot the evolution of angles
    describing the orientation of our fiducial NS-BH and BBH binaries versus
    time to coalescence.  For both the NS-BH (top left) and BBH (top right)
    binaries, the radiation-frame angle $\iota$ evolves significantly, while
    the system-frame angles $\psij,\theta_{JL},\theta_{JN}$ evolve slowly.
    For the single-spin NS-BH binary the spin maintains a constant tilt
    relative to $\hat{L}$ (not shown).  For the double-spin BBH binary, the
    tilts of the component spins relative to $\hat{L}$ do vary by a few tenths
    of a radian (lower left panel).  In the lower right panel we plot the
    azimuthal angles $\phi_{JL}$ describing the motion of $\hat{L}$ moving in
    its cone about $\hat{J}$, and $\phi_{12}$, the azimuthal separation of
    component spins measured relative to $\hat{L}$ for the BBH binary. 
 \label{fig:NSBH_angles}
\label{fig:BBH_angles_1}
}
\end{figure*}

We note that the simple precession model presented in~\cite{Apostolatos:1994mx}
primarily focused on the special cases of equal masses and/or single spin
binaries.  This was so the precession
equations~(\ref{eq:dLdt})-(\ref{eq:dS2dt}) would imply that $\mathbf{S}_1 \cdot
\mathbf{S}_2 = {\rm const.}$ and simplify the behavior.  However, even for a
generic case with unequal masses and spins, the simple precession model is
still a good \emph{qualitative} description of the precessing binary's
behavior.  One difference is that the orbital angular momentum exhibits
nutation.  That is, $\hat{L}$ bobs up and down as it moves on its cone about
$\hat{J}$.  The cone opening angle still tends to grow as orbital angular
momentum is radiated away, but it is no longer a monotonic increase.  Compare
the single spin NS-BH binary to the double spin BBH binary in
Fig.~(\ref{fig:BBH_angles_1}).  The angle between $\hat{J}$ and $\hat{L}$,
$\thetajl$, nutates strongly in the latter but not in the former.  Note however
that the angles $\thetajn$ and $\psij$, the inclination and polarization angles
of the total angular momentum, respectively, are nearly constant in each case,
as $\hat{J}$ remains essentially fixed.  Additionally,
Fig.~(\ref{fig:BBH_angles_1}) shows the tilt of the component spins and the
total spin $\mathbf{S} = \mathbf{S}_1 + \mathbf{S}_2$ relative to $\hat{L}$.
Note that these angles can change by a few tenths of a radian (with the total
spin oscillating less than the individual spins), while for the single-spin
case (not shown) $\hat{L} \cdot \hat{S}_1$ is constant throughout the
evolution.

\subsection{Previous coordinate conventions}
\label{subsec:prev_coords}

As we have summarized, computing the orbital dynamics of a precessing
post-Newtonian waveform involves numerically integrating the coupled set of
ordinary differential equations (ODEs)
in~(\ref{eq:energy_balance})-(\ref{eq:dS2dt}).  Typically, one specifies a
binary with parameters $m_1$ and $m_2$, plus the initial Cartesian components
of the spin and orbital angular momentum vectors specified at the minimum
frequency at which the signal enters the detector's sensitive band. This is
indeed an obvious, natural choice to specify the initial conditions of the ODEs
and is what has previously been used to generate CBC waveforms in the LAL
software.

Actually, for precessing binaries LAL codes adopt a simplifying convention,
without loss of generality, so that the orbital angular momentum vector can be
specified by a single inclination angle, $\iota$. In particular, the $z$-axis
of the Cartesian frame is taken to be the direction of propagation of the
gravitational wave. The orbital angular momentum is assumed to lie in the
$x$-$z$ plane, such that $\hat{L} = (\sin \iota, 0, \cos \iota)$,  and hence
such that the projection of $\hat{L}$ on the plane of the sky is in the
direction $(\cos \psil,\sin \psil) = (1,0)$.  Therefore, the LAL waveform
generation routines depend on the parameters
\begin{equation} \label{eq:radiation_frame} 
    \{m_1, m_2, \iota, (\mathbf{S}_{1x}, \mathbf{S}_{1y}, \mathbf{S}_{1z}), (\mathbf{S}_{2x}, \mathbf{S}_{2y}, \mathbf{S}_{2z})\}\ .
\end{equation}
This Cartesian parameterization, with the $z$-axis along the direction of GW
propagation, is commonly called the \emph{radiation frame}.  Previous parameter
estimation efforts, such
as~\cite{2008ApJ...688L..61V,2009CQGra..26k4007R,2014arXiv1403.0544O,Aasi:2013jjl},
would estimate these radiation frame parameters,\footnote{Additionally, the
    time and phase of some reference point $t_{\rm ref}$, $\phi_{\rm ref}$, sky
    location $(\delta, \alpha)$, luminosity distance $D_L$ and polarization
    angle $\psil$ are needed, but they merely describe the orientation of the
    binary relative to the observer and do not affect the orbital dynamics in
    any way.}

While the radiation frame is very convenient for generating waveforms, it has
several drawbacks for parameter estimation.  First of all, the Cartesian
angular momentum coordinates of a binary, and hence their projection on the
plane of the sky (i.e., $\psil$), can vary considerably over its evolution.
Fig.~(\ref{fig:NSBH_angles}) shows how the inclination of the orbital plane
($\iota$) varies across the observed signal, and the Cartesian components of
$\hat{S}_i$ (not shown) cover most of their allowed range $[-1,1]$.  Otherwise
identical binaries that happen to be at two different points along the
precession cone at the reference point will have very different Cartesian
components.  Thus, systems that produce similar GW signals are spread across
the parameter space in complex ways, and parameter estimation analyses must map
out these complicated correlations.  Furthermore, the Cartesian components are
specified at the low frequency limit, where detectors have poor sensitivity.
Two binaries with similar vector components at $f_{\rm min}$ might have rather
different component values when the signal is in the sensitive band of the
detector.
 
\subsection{Nearly-conserved coordinates}
\label{subsec:cons_coords}

Motivated by the simple precession picture of Apostolatos~et.~al., we
parameterize the binary configuration via a set of angles that describe the
position and shape of the precession cone, as well as where $\hat{L}$ is along
its cone. In particular, we parameterize a binary configuration with
\begin{equation} \label{eq:system_frame} \{ m_1, m_2, \chi_1, \chi_2, \thetajn,
    \tilta, \tiltb, \phiab, \phijl  \}\ .  \end{equation} Here $0 \leq
    \chi_{1,2} \leq 1$ are the spin magnitudes, $\thetajn$ is the inclination
    between the total angular momentum and the direction of propagation,
    $\theta_{LS_{1,2}}$ are the inclinations of each spin relative to $\hat{L}$
    (commonly referred to as tilts), $\phi_{12}$ is the azimuthal angle of
    $\hat{S}_2 - \hat{S}_1$ measured relative to $\hat{L}$\footnote{Only the
        relative azimuthal difference between the spin vectors matters. The
        absolute azimuth, relative to, say $\hat{N}$, is implicitly set by
    $\phijl$.}, and $\phijl$ is the azimuthal position of $\hat{L}$ on its cone
    about $\hat{J}$ (with $\hat{N}$ setting the zero of azimuth).  We refer to
    this parameterization as precessing system coordinates (or the ``system
    frame'' for short). 

We note that this parameterization, like the standard Cartesian one, requires 9
parameters to specify the orbital dynamics. The system frame parameterization
captures all degrees of freedom in the binary but does not overdetermine it.
Given values for all the parameters of either the system or radiation frame,
one can compute the values of the other parameterization through a series of
Euler rotations. We have implemented such transformations in the LAL software.
As a practical implementation, parameter estimation routines propose values for
the parameters of Eq.~\ref{eq:system_frame}, these are transformed into the
parameters of Eq.~\ref{eq:radiation_frame} and passed to waveform generation
routines which take input in terms of these parameters.

An advantage of the precessing system frame is that many of the
parameters are nearly conserved for the duration that the signal is in
band. It is well known the masses and spin magnitudes do not change
significantly during inspiral, and Apostolatos~et.~al. showed that
$\thetajn$ is very nearly constant. For the case of a single-spin
binary, Apostolatos~et.~al. showed $\theta_{LS_1}$ is constant at
leading order. For a double-spin binary, the angles $\theta_{LS_1}$,
$\theta_{LS_2}$, $\phi_{12}$ need not be conserved -- indeed
Fig.~\ref{fig:BBH_angles_1} shows the first two vary by a few tenths
of a radian, while $\phi_{12}$ grows through $\mysim$two full cycles.
However, it should be noted that the BBH binary plotted in
Fig.~\ref{fig:BBH_angles_1} was intentionally chosen as an extreme
case.  Many double spin binaries will have significantly less
variation in these angles.  For example, there are known to be
\emph{spin resonances} where both spins and $\hat{L}$ get locked into
a coplanar configuration~\cite{Schnittman:2004vq}.  Lastly, $\phijl$
is not constant, but does increase steadily and monotonically,
essentially chirping on a precessional timescale.  As we will see in
Section~\ref{sec:results}, using these nearly-conserved coordinates
improve the convergence rates of Bayesian parameter estimation.

\subsection{In-band coordinates}
\label{subsec:inband_coords}

Precessional motion causes amplitude and phase modulations, and can change the
polarization content of the observed waveform (i.e. the relative strength of
$h_+$ and $h_\times$). For example, we see from
Eqs.~(\ref{eq:plus_pol})--(\ref{eq:cross_pol}) that $h_+$ and $h_\times$
polarizations will have equal strength when the binary is face on, but
$h_\times$ vanishes when the binary is edge on.  The detectors will
essentially be measuring the polarization content when the signal is in the
``bucket'', or the most sensitive frequency band of the detector. They will
necessarily be less sensitive when the binary is at lower frequencies near the
seismic cutoff.  Therefore, it stands to reason that it is more difficult to
constrain the orientation of the binary at low frequencies, where there is
little sensitivity, than at frequencies of peak sensitivity, which are
typically $\sim 100$--$200$ Hz for ground based detectors.

It is therefore unfortunate that waveform generation routines and parameter
estimation jump proposals in LAL have until now specified initial conditions
for binary orientation at the low frequency limit, where the sensitivity is
worst. If the orientations of two binaries are similar at low frequencies, they
need not be similar in the bucket. For example, if the masses and/or spin
magnitudes are a bit different, they may precess at different rates and could
be at different points along similar precession cones when in bucket, thus
having very different polarization content for certain observers. Or, they
could move along very dissimilar precession cones that happen to be nearly
tangent at a certain point at low frequency.

Fortunately, we note that the differential equations needed to evolve the
orbital dynamics, Eqs.~(\ref{eq:energy_balance})--(\ref{eq:dS2dt}), can be
integrated backwards in time just as easily as forwards. The same is true for
any post-Newtonian waveform model. As a practical implementation, we choose
some reference frequency, $f_{\rm ref}$, and specify the ``initial'' conditions
for the binary orientation at that gravitational-wave frequency.  We then make
two calls to evolve the orbital dynamics: one integrates
Eqs.~(\ref{eq:energy_balance})--(\ref{eq:dS2dt}) backwards in time until it
reaches the minimum frequency of detector sensitivity $f_{\rm min}$, and the
other integrates forward in time to a frequency $f_{\rm end}$, which can be the
minimum energy circular orbit (MECO) or some other waveform stopping condition.
We then stitch together the waveform time series from each integration to get a
seamless waveform that covers the full frequency range $[ f_{\rm min}, f_{\rm
end} ]$.  As we have implemented it, this method of two-way integration agrees
with the standard approach of forward integration from $f_{\rm min}$ to within
numerical precision, and there is virtually no difference in the speed of
waveform generation.

In principal, we can choose $f_{\rm ref}$ to be any value which the binary
reaches before the MECO or other termination condition. However, as expected,
we find in Section~\ref{sec:results} that parameter estimation codes are most
efficient when choosing $f_{\rm ref}$ near where the detector has peak
sensitivity.

\begin{table*}
    \begin{tabular}{c|ccccccccc|cc|c}
        Binary Type&$m_1$&$m_2$& $\chi_1$ & $\chi_2$ & $\thetajn$ & $\tilta$ & $\tiltb$ &  $\phiab$ 
            & $\phijl$ & $\thetajl$ & $\psij$ & $\rho_{\rm net}$ \\ 
        \hline
        BNS & 1.2--1.6 & 1.2--1.6 & 0--0.05 & 0--0.05 & -- & -- & -- & -- & -- & -- & -- & 10.3--33.4 \\
        NS-BH & 10 & 1.4 & 1 & 0 & 1.22  & 1.175 &  N/A & N/A & $\pi/4$ &$\pi/4$  & 2.36 & 20.3  \\ 
        BBH & 8 &5 & 0.8 & 0.9 & $\pi/6$ & $\pi/2$ & $\pi/2$ & $\pi/2$ & 0 & 0.36 &  1.57  & 19.1 \\
    \end{tabular}
    \caption{\label{table1} The precessing system parameters (see
        Eq.~(\ref{eq:system_frame}) and the surrounding text) for our BNS
        population and fiducial NS-BH and BBH binaries.  For the NS-BH and BBH
        systems we provide the opening angle of the precession cone
        ($\theta_{JL}$), the polarization angle of the total angular momentum
        ($\psij$), and the network SNR ($\rho_{\rm net}$), for a
        three-detector network of early aLIGO-AdV detectors.  A population of
        15 BNS systems with masses and spins chosen uniformly from the
        specified ranges, and orientations chosen isotropically, and the range of
        their SNRs in two-detector early aLIGO detectors are shown.}
\end{table*}

\begin{figure*}
    \includegraphics[width=0.32\textwidth]{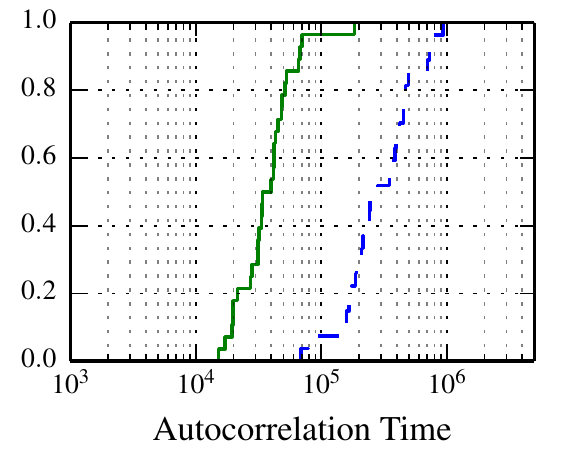}
    \includegraphics[width=0.32\textwidth]{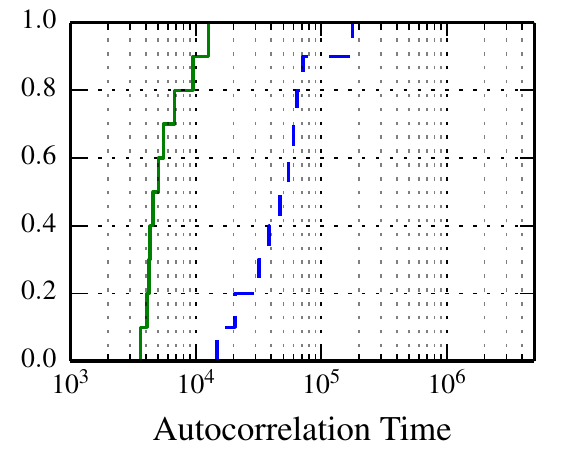}
    \includegraphics[width=0.32\textwidth]{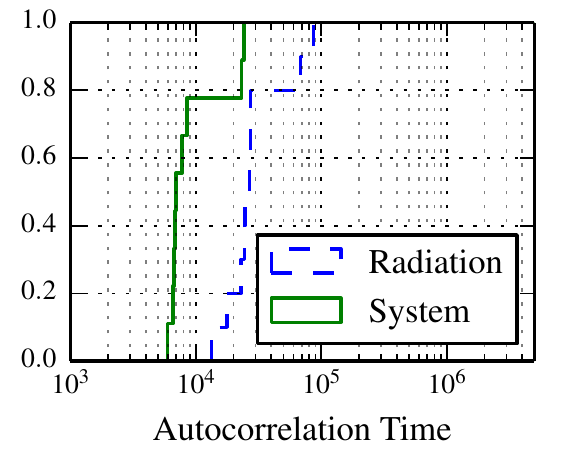}
    \caption{\textbf{Autocorrelation times}: Cumulative histogram of the largest
    intrinsic-parameter ACT of each MCMC chain, for signals analyzed using the
    radiation frame (with parameters defined at 40 Hz) and precessing system frame
    (parameters defined at 100 Hz).
    \emph{Left panel}\label{fig:BNS_ACT} Results
    for the 15 BNS signals; using the system frame reduces ACTs by a median factor
    of \bnsfactor.
    \emph{Center panel}: \label{fig:NSBH_acls} Results for each analysis of the
    NS-BH binary described in Table~\ref{table1}.  The system frame is found to
    have a median improvement of a factor of \nsbhfactor~in efficiency over the
    radiation frame.
    \emph{Right panel}: \label{fig:BBH_acls} Results for each analysis of the  BBH
    binary described in Table~\ref{table1}.  The system frame is found to have a
    median improvement of a factor of \bbhfactor~in efficiency over the radiation
    frame.}
\end{figure*}

\section{Parameter estimation with Markov Chain Monte Carlo}
\label{sec:MCMC}

For CBC parameter estimation, Markov chain Monte Carlo (MCMC) methods are used
to sample the full 15 dimensional posterior distribution as a function of the
parameters describing the circularized compact binary merger.  These methods
employ serial Markov chains that stochastically wander the parameter space
through the use of various proposal distributions.  By accepting or rejecting
proposed jumps according to the Metropolis-Hastings
ratio~\cite{metropolis53,hastings}, samples recorded by the Markov chains are
distributed with a density proportional to the target posterior probability
density.  To ensure each sample is an independent draw from the posterior
distribution, samples are first thinned based on the correlations present in
the chain.  Thus, assuming the likelihood function is equally expensive to
compute at all times, the efficiency of the MCMC sampler will ultimately be
decided by the maximum 1-D autocorrelation time (ACT), $t_{\rm max}$, of the
chains.  We estimate the 1-D autocorrelation time $t$ for random variable $X$
as the smallest $s$ that satisfies
\begin{equation} \label{eqn:act}
    1 + \frac{2}{C_0}\sum_{\tau=1}^{M s} C(\tau) < s,
\end{equation}
where $C(\tau)$ is the autocorrelation function $C(\tau) =
E[(X_t-\mu)(X_{t+\tau}-\mu)]/\sigma^2$, with $\mu$ and $\sigma$ being the mean
and standard deviation of $X$, respectively, and $C_0=C(\tau=0)$ is the
zero-lag autocorrelation.  $M$ is a tunable parameter, ensuring the stability
of the ACT estimate by requiring the length of the window used to estimate the
ACT to be at least $M$ times the estimated ACT.  We have empirically found $M=5$
to produce reliable ACT estimates.

In order to minimize ACTs, it is critical that the acceptance rates of jump
proposals are balanced with the correlations they introduce.  In general, the
most efficient proposal distribution is the target distribution.  The target
distribution is typically unknown, and a collection of generally useful
proposal distributions are used in its place.  The most basic jump proposal
typically used for MCMC sampling is a local Gaussian, centered on the current
location of the chain.  The width of this Gaussian in each dimension will
affect the acceptance rates and ACTs.  In the small-width limit jump acceptance
rates approach 1, however the very small steps of the chain greatly increase
ACTs.  In the large-width limit jump acceptance rates approach 0, also
resulting in large ACTs due to many repeated samples in the chains.  In the
idealized case of an $N$-dimensional Gaussian target distribution composed of
$N$ independent 1-dimensional Gaussians, it can be shown that the ideal
acceptance rate of this proposal that minimizes ACTs is
$\mysim23.4$\%~\cite{Gelman96}. In the case of non-Gaussian target
distributions, this acceptance rate is not necessarily optimal, and this
proposal (even with optimized widths) can be very inefficient.

For CBC parameter estimation, the local Gaussian proposal distribution, though
not the only one, is the proposal used most often.  Since this proposal is
optimal for Gaussian target distributions, the parameterization used for
sampling should be chosen such that the posterior is as close to Gaussian as
possible.  For the purpose of determining the efficiency gains from the system
frame, we will focus on the sampling of the intrinsic parameters: $\left\{ \mc,
\q, \iota, \chi_1, \thetaa, \phia, \chi_2, \thetab, \phib\right\}$ for the
radiation frame and $\left\{ \mc, \q, \thetajn, \phijl, \chi_1, \tilta, \chi_2,
\tiltb, \phiab\right\}$ for the system frame, where ${\cal M} = (m_1
m_2)^{3/5}(m_1+m_2)^{-1/5}$ is the chirp mass and $\q$ is the asymmetric mass
ratio $m_2/m_1$ defined such that $0 < q \leq 1$.

\section{Results}
\label{sec:results}

To assess the overall improvement in MCMC sampling efficiency, we have
simulated GW inspiral signals from the three main types of compact binaries
expected to be observed by advanced ground-based detectors: binary neutron
stars (BNS), neutron star-black hole systems (NS-BH), and binary black hole
systems (BBH). The range of parameters studied are provided in Table
\ref{table1}.  For BNS systems we focus on the impact of the system frame on
analyses during the first year of the advanced-detector era (2015), assuming
Gaussian noise from the Hanford and Livingston detectors with ``early'' aLIGO
sensitivity, as defined in ~\cite{Aasi:2013wya}.

For the fiducial NS-BH and BBH systems we move to the projected 2016
three-detector network, with ``mid'' aLIGO sensitivity for Hanford and
Livingston and ``early'' AdV sensitivity for Virgo, as defined
in~\cite{Aasi:2013wya}.  For these systems we also consider the choice of
reference frequency, and its impact on parameter constraints and sampling
efficiency.

\subsection{Binary Neutron Stars}
\label{subsec:bns_results} With Virgo not coming online until 2016, the first
year of the advanced-detector era will only see two operational detectors.
Since binary neutron stars are observationally confirmed sources that aLIGO is
expected to be sensitive to, we place particular emphasis on the impact of
system frame analyses on BNS signals in the early advanced-detector era.  To
this end, we have randomly selected 15 BNS signals detected from an
astrophysically distributed set of injections in a simulated two-detector
network.  Motivated by the observed neutron star population to date, the
injection population was chosen to have dimensionless spin magnitudes
distributed uniformly below $0.05$~\cite{2005yCat.7245....0M}, with isotropic
spin orientations, and component masses drawn uniformly between 1.2 and 1.6
$\text{M}_\odot$.  This injection set was distributed uniformly in the local
universe, and 15 randomly selected signals detected by
$\gstlal$~\cite{2012ApJ...748..136C} were chosen for the purposes of this
study~\cite{first2years}.

\begin{figure*}
    \includegraphics[width=0.45\textwidth]{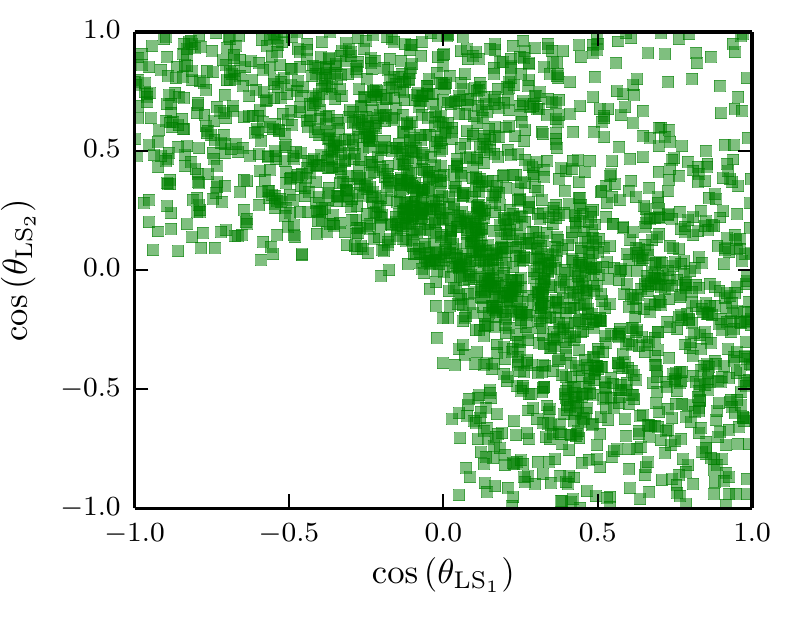}
    \includegraphics[width=0.45\textwidth]{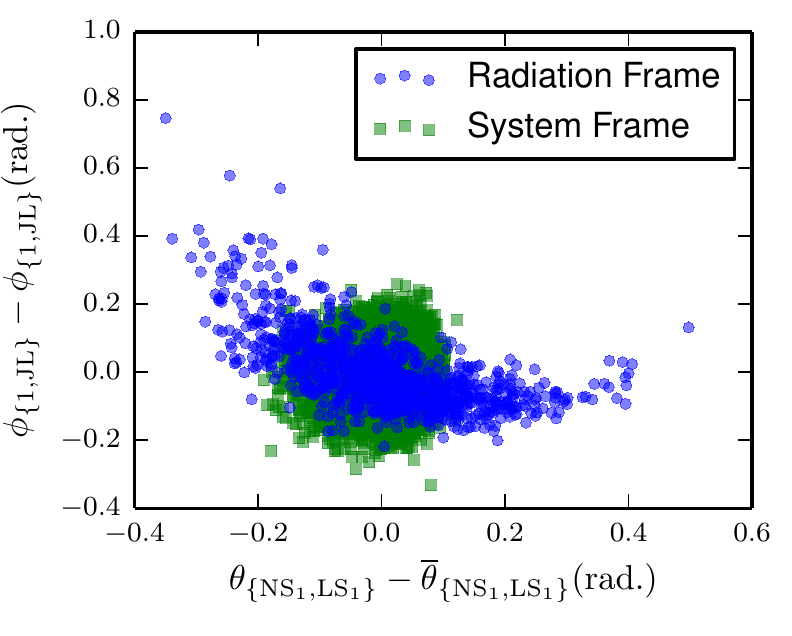}
    \caption{\emph{Left panel} \label{fig:BNS_tilts} MCMC samples from the
    posterior of a simulated BNS signal.  The region where both spins have a
    negative $\hat{L_z}$ component is excluded.  This information is easily
    described, and sampled, in the system frame.  \emph{Right panel}
    \label{fig:NSBH_tilts} Posterior samples from the analyses of an NS-BH
    injection, centered on the mean. $\theta$ and $\phi$ correspond to
    $\thetaa$ and $\phia$ in the radiation frame, and $\tilta$ and $\phijl$ in
    the system frame.  The strong, non-linear correlation between spin
    parameters in the radiation frame is difficult to sample, resulting in
    large ACTs.  The system frame parameterization eliminates this correlation,
    allowing for more efficient MCMC sampling.}
\end{figure*}

We compare the maximum one-dimensional ACT from each MCMC chain sampling in the
radiation frame ($f_{\rm ref}=40$Hz) and system frame ($f_{\rm ref}=100$Hz)
parameterizations. We find that the precessing system frame gives ACTs which
are shorter by $\bnsfactor$, initially surprising due to the minimal effects of
spin on the GW signal.  Fig.~\ref{fig:BNS_ACT} shows the cumulative
distribution of ACTs for the intrinsic parameters of the relevant frame across
the 15 BNS systems.  Even though these systems have very low spin, and
therefore very little precession, a long-known degeneracy between mass ratio
and spin magnitude~\cite{Baird:2012cu,2013ApJ...766L..14H,2014arXiv1401.0939T}
comes into play.  In the simplified picture of a BNS system with spins aligned
with the orbital angular momentum, changes to the waveform from increasing spin
magnitude cannot be distinguished from increases in mass ratio, and visa versa.
Because BNS systems are very close to equal mass, the posterior is highly
skewed towards lower mass ratios, and thus higher spin magnitudes in the
direction of the angular momentum.  Support for anti-aligned spin would require
support for higher mass ratios, which would be unphysical.  Parametrically this
is easily described in the system frame as $\text{cos}\left(\tilta\right),
\text{cos}\left(\tiltb\right) > 0$, resulting in an excluded region in
$\tilta$-$\tiltb$ space as seen in Fig.~\ref{fig:BNS_tilts}.  In the radiation
frame however, this region of parameter space is non-trivially defined as a
function of all four spin orientation angles, which no proposal used by the
MCMC is able to navigate efficiently.

\subsection{Neutron Star-Black Hole and Binary Black Hole Systems}
\label{subsec:fiducial_results}

\begin{figure}
    \includegraphics[width=0.45\textwidth]{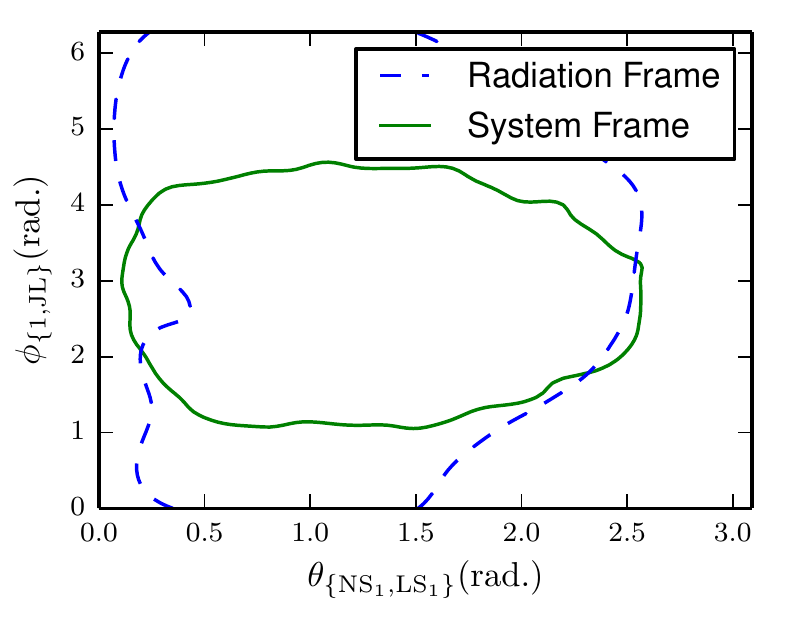}
    \caption{\label{fig:BBH_tilts} 95\% posterior credible regions from analyses of
        a BBH injection.  $\theta$ and $\phi$ correspond to $\thetaa$ and
        $\phia$ in the radiation frame, and $\tilta$ and $\phijl$ in the system
        frame. The posterior in the radiation frame parameterization has
        significant structure.  In the system frame parameters, support is
        largely confined to a single uncorrelated mode that crosses the cyclic
        boundary of $\phijl$, proving much more efficient to sample.}
\end{figure}

\begin{figure*}
    \includegraphics[width=0.45\textwidth]{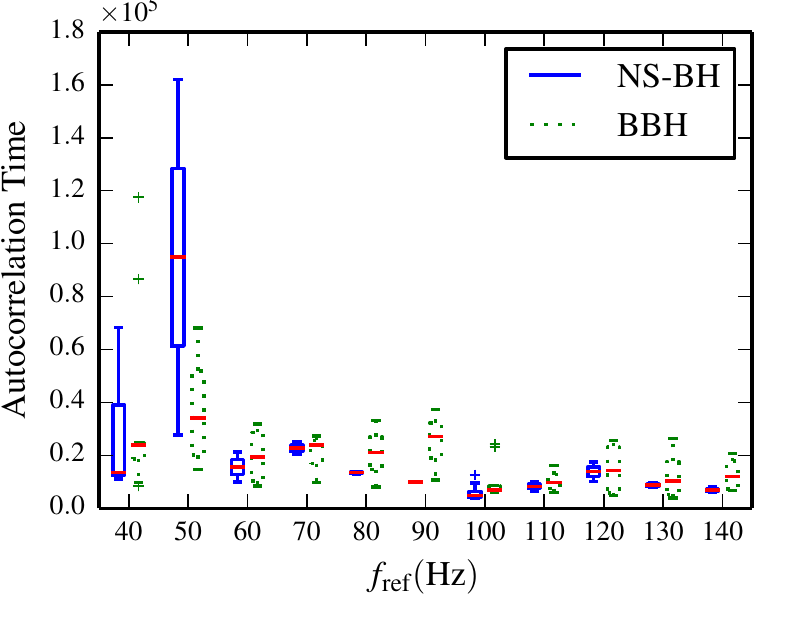}
    \includegraphics[width=0.45\textwidth]{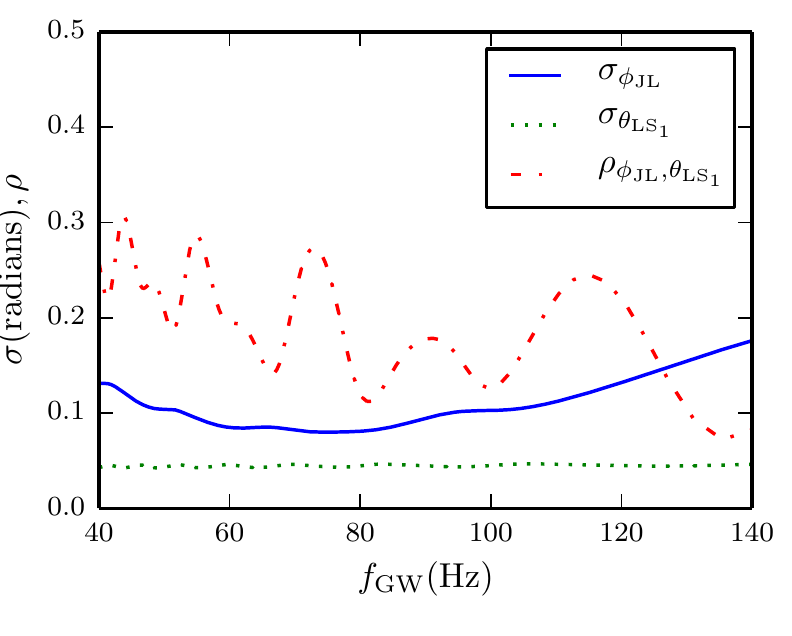}

    \caption{
    \emph{Left panel}: \label{fig:frefACLs} Box-and-whisker plots summarizing the
    distributions of maximum intrinsic-parameter ACTs for the fiducial NS-BH
    and BBH systems for the eleven tested reference frequencies.  Boxes
    indicate the first and third quartiles, with interior lines indicating
    medians, whiskers showing the range of data within 1.5 inner-quartile-widths,
    and `+'s indicating points outside this range.
    \emph{Right panel}: \label{fig:NSBH_std} The standard deviations and correlation of
    the $\tilta$ and $\phijl$ distributions of the NSBH fiducial system over
    the evolution of the waveform.  $\tilta$ is found to have a relatively
    stationary distribution, while $\phijl$ is best constrained at $\mysim70$
    Hz.  The correlation between parameters, though highly variable, has a
    minimum near 140 Hz.}
\end{figure*}

\subsubsection{System Frame Efficiency}
Based on current estimates from low-mass x-ray binaries, most black holes in
binary systems are believed to have significant spin angular
momentum~\cite{McClintock:2011zq}.  Thus, even slight mis-alignments for
NS-BH and BBH systems will lead to precession of the orbital plane, and
modulation of the measured gravitational wave signal.  For the case of NS-BH
systems, the neutron star has negligible angular momentum compared to the black
hole, and the system will undergo simple precession, as described in
Section~\ref{subsec:precessing_behavior}.  For BBH systems, both components will
have comparable angular momenta.  In this regime the precession behavior
becomes more complex, with $\hat{\mathbf{L}}$ nutating as it moves along its
cone and the spins not maintaining a fixed orientation relative to
$\hat{\mathbf{L}}$.  We have chosen two fiducial NS-BH and BBH binaries with
parameters given in Table~\ref{table1} to assess the importance of well-chosen
parameters for strongly precessing binaries.

For the NS-BH system, the spin of the BH is well constrained.  In the radiation
frame, shown in Fig.~\ref{fig:NSBH_tilts}, the BH spin is constrained to a
single mode with strong non-linear correlation between $\thetaa$ and $\phia$.
In the system frame, this constraint is mainly in $\tilta$ and $\phijl$, but
with little correlation between these parameters.  Sampling correlations like
that found in the radiation frame proves very inefficient, since the jump
proposal used most often proposes jumps in one dimension at a time.

Both components of the BBH system in this study have significant angular
momentum, making degeneracies stronger between them.  For this particular
system, the constraints placed on the spin parameters are much weaker than for
the NS-BH system's primary spin.  However, the system frame still isolates
physical features in the waveform, and reduces the correlation between spinning
parameters.  Figure~\ref{fig:BBH_tilts} shows the 95\% credible regions in the
primary spin's orientation parameters.  In the radiation frame, the posterior
is highly structured, with regions of high correlation between parameters that
make MCMC sampling inefficient.  In the system frame however, posterior support
is confined to a single uncorrelated mode that stretches across the cyclic
boundary of $\phijl$, ideal for sampling efficiently with the proposal set
employed.

To assess the improved sampling efficiency using the system frame, MCMC chains
were used to analyze the fiducial NS-BH and BBH binaries described in
Table~\ref{table1}. In Fig.~(\ref{fig:NSBH_acls}) we
show the cumulative distributions of one-dimension ACTs for the intrinsic
parameters in the parameterization relevant to the frame.

\subsubsection{Reference Frequency}

It is clear the system frame parameterization, with evolving parameters
defined at 100 Hz, is beneficial for MCMC sampling.  To justify this choice in
reference frequency, we have conducted a suite of analyses where the system
frame parameters are specified at differing reference frequencies.
Figure \ref{fig:frefACLs} shows the efficiency of MCMC analyses for eleven
different different choices of reference frequency.  We find that a reference
frequency near 100 Hz achieves the most efficient MCMC sampling.

By taking the posterior samples from an analysis with $f_{\rm ref}=40$ Hz, we
can use the PN equations outlined in Section~\ref{sec:coords} to evolve the
posterior samples to later times in the inspiral.  We have done this for the
NS-BH system to determine how the primary component's spin orientation is
constrained over the course of the inspiral.  Figure \ref{fig:NSBH_std} shows
the constraint on $\tilta$ is roughly uniform across the waveform, as we would
expect under the simple precession evolution.  The constraints on $\phijl$,
however, vary significantly over the inspiral with a minimum around $\mysim70$
Hz.  The correlation between $\tilta$ and $\phijl$ has a less consistent
evolution, but for this system has a minimum at $\mysim140$ Hz.  Since the jump
sizes for the Gaussian proposal are adapted to the width of the posterior in
each dimension separately, the correlation is expected to impact the sampling
efficiency more than the one-dimensional standard deviations.  This is
reflected by the ACTs in the left panel of Fig.~\ref{fig:frefACLs}, where the
ACTs can be seen to oscillate roughly in accordance with the correlation in the
right panel.  This picture is likely to change for different systems, which is
reflected by the BBH ACTs.  Ultimately we find that choosing a reference
frequency of $\mysim100$ Hz should reliably result in efficient sampling.

\section{Conclusions}
\label{sec:conclusions}

In this work we present an improved method for parameterizing precessing
compact binary coalescence waveforms, which we have implemented in the LAL
software library.  We have compared the efficiency of MCMC sampling with both
parameterizations for BNS, NS-BH, and BBH systems, and find improvements of
factors of \bnsfactor, \nsbhfactor, and \bbhfactor, respectively.  We have
also determined the sampling efficiency and parameter constraints for system
frame parameters for a range of reference frequencies.  We find that using
a reference frequency of $\mysim100$ Hz will ensure efficient sampling for
both NS-BH and BBH signals.

While we assessed the improvement using a specific parameter estimation
algorithm (\mcmc) and waveform model (a time domain SpinTaylorT4
implementation), this method can be expected to improve the performance of any
underlying sampling method (e.g., nested sampling) and any precessing waveform
implementation  (e.g., frequency-domain templates~\cite{2014arXiv1404.3180C},
precessing effective-one-body~\cite{2014PhRvD..89h4006P}, et cetera).  By
moving to a more astrophysically intuitive parameterization, physical features
in the waveform, such as a lack of precession, are described using fewer
parameters.  This reduces the correlation between model parameters, and
increases the efficiency of parameter estimation analyses.  By additionally
specifying the orientation of the binary and its components at a point near the
detectors' peak sensitivity, the efficiency or parameter estimation can be
further increased.

\begin{acknowledgments}
BF was supported by NSF fellowship DGE-0824162, with computing resources funded
by PHY-1126812.
EO and ROS would like to acknowledge support from NSF grants
PHY-0970074 and PHY-1307429. 
\end{acknowledgments}

\bibliography{references}

\begin{thebibliography}{50}
\expandafter\ifx\csname natexlab\endcsname\relax\def\natexlab#1{#1}\fi
\expandafter\ifx\csname bibnamefont\endcsname\relax
  \def\bibnamefont#1{#1}\fi
\expandafter\ifx\csname bibfnamefont\endcsname\relax
  \def\bibfnamefont#1{#1}\fi
\expandafter\ifx\csname citenamefont\endcsname\relax
  \def\citenamefont#1{#1}\fi
\expandafter\ifx\csname url\endcsname\relax
  \def\url#1{\texttt{#1}}\fi
\expandafter\ifx\csname urlprefix\endcsname\relax\def\urlprefix{URL }\fi
\providecommand{\bibinfo}[2]{#2}
\providecommand{\eprint}[2][]{\url{#2}}

\bibitem[{\citenamefont{Harry et~al.}(2010)}]{Harry:2010zz}
\bibinfo{author}{\bibfnamefont{G.~M.} \bibnamefont{Harry}}
  \bibnamefont{et~al.}, \bibinfo{journal}{Class. Quant. Grav.}
  \textbf{\bibinfo{volume}{27}}, \bibinfo{pages}{084006}
  (\bibinfo{year}{2010}).

\bibitem[{\citenamefont{Acernese et~al.}(2009)}]{aVirgo}
\bibinfo{author}{\bibfnamefont{F.}~\bibnamefont{Acernese}} \bibnamefont{et~al.}
  (\bibinfo{year}{2009}), \bibinfo{note}{{Virgo Technical Report 0027A-09}}.

\bibitem[{\citenamefont{Abadie et~al.}(2010)}]{Abadie:2010cf}
\bibinfo{author}{\bibfnamefont{J.}~\bibnamefont{Abadie}} \bibnamefont{et~al.}
  (\bibinfo{collaboration}{LIGO Scientific Collaboration, Virgo
  Collaboration}), \bibinfo{journal}{Class.Quant.Grav.}
  \textbf{\bibinfo{volume}{27}}, \bibinfo{pages}{173001}
  (\bibinfo{year}{2010}), \eprint{1003.2480}.

\bibitem[{\citenamefont{McClintock et~al.}(2011)\citenamefont{McClintock,
  Narayan, Davis, Gou, Kulkarni et~al.}}]{McClintock:2011zq}
\bibinfo{author}{\bibfnamefont{J.~E.} \bibnamefont{McClintock}},
  \bibinfo{author}{\bibfnamefont{R.}~\bibnamefont{Narayan}},
  \bibinfo{author}{\bibfnamefont{S.~W.} \bibnamefont{Davis}},
  \bibinfo{author}{\bibfnamefont{L.}~\bibnamefont{Gou}},
  \bibinfo{author}{\bibfnamefont{A.}~\bibnamefont{Kulkarni}},
  \bibnamefont{et~al.}, \bibinfo{journal}{Class.Quant.Grav.}
  \textbf{\bibinfo{volume}{28}}, \bibinfo{pages}{114009}
  (\bibinfo{year}{2011}), \eprint{1101.0811}.

\bibitem[{\citenamefont{Kidder}(1995)}]{Kidder:1995zr}
\bibinfo{author}{\bibfnamefont{L.~E.} \bibnamefont{Kidder}},
  \bibinfo{journal}{Phys.Rev.} \textbf{\bibinfo{volume}{D52}},
  \bibinfo{pages}{821} (\bibinfo{year}{1995}), \eprint{gr-qc/9506022}.

\bibitem[{\citenamefont{Arun et~al.}(2009)\citenamefont{Arun, Buonanno, Faye,
  and Ochsner}}]{Arun:2008kb}
\bibinfo{author}{\bibfnamefont{K.~G.} \bibnamefont{Arun}},
  \bibinfo{author}{\bibfnamefont{A.}~\bibnamefont{Buonanno}},
  \bibinfo{author}{\bibfnamefont{G.}~\bibnamefont{Faye}}, \bibnamefont{and}
  \bibinfo{author}{\bibfnamefont{E.}~\bibnamefont{Ochsner}},
  \bibinfo{journal}{Phys. Rev.} \textbf{\bibinfo{volume}{D79}},
  \bibinfo{pages}{104023} (\bibinfo{year}{2009}), \eprint{0810.5336}.

\bibitem[{\citenamefont{Apostolatos et~al.}(1994)\citenamefont{Apostolatos,
  Cutler, Sussman, and Thorne}}]{Apostolatos:1994mx}
\bibinfo{author}{\bibfnamefont{T.~A.} \bibnamefont{Apostolatos}},
  \bibinfo{author}{\bibfnamefont{C.}~\bibnamefont{Cutler}},
  \bibinfo{author}{\bibfnamefont{G.~J.} \bibnamefont{Sussman}},
  \bibnamefont{and} \bibinfo{author}{\bibfnamefont{K.~S.}
  \bibnamefont{Thorne}}, \bibinfo{journal}{Phys. Rev.}
  \textbf{\bibinfo{volume}{D49}}, \bibinfo{pages}{6274} (\bibinfo{year}{1994}).

\bibitem[{\citenamefont{Kidder et~al.}(1993)\citenamefont{Kidder, Will, and
  Wiseman}}]{Kidder:1992fr}
\bibinfo{author}{\bibfnamefont{L.~E.} \bibnamefont{Kidder}},
  \bibinfo{author}{\bibfnamefont{C.~M.} \bibnamefont{Will}}, \bibnamefont{and}
  \bibinfo{author}{\bibfnamefont{A.~G.} \bibnamefont{Wiseman}},
  \bibinfo{journal}{Phys.Rev.} \textbf{\bibinfo{volume}{D47}},
  \bibinfo{pages}{4183} (\bibinfo{year}{1993}), \eprint{gr-qc/9211025}.

\bibitem[{\citenamefont{Faye et~al.}(2006)\citenamefont{Faye, Blanchet, and
  Buonanno}}]{Faye:2006gx}
\bibinfo{author}{\bibfnamefont{G.}~\bibnamefont{Faye}},
  \bibinfo{author}{\bibfnamefont{L.}~\bibnamefont{Blanchet}}, \bibnamefont{and}
  \bibinfo{author}{\bibfnamefont{A.}~\bibnamefont{Buonanno}},
  \bibinfo{journal}{Phys.Rev.} \textbf{\bibinfo{volume}{D74}},
  \bibinfo{pages}{104033} (\bibinfo{year}{2006}), \eprint{gr-qc/0605139}.

\bibitem[{\citenamefont{Blanchet et~al.}(2006)\citenamefont{Blanchet, Buonanno,
  and Faye}}]{Blanchet:2006gy}
\bibinfo{author}{\bibfnamefont{L.}~\bibnamefont{Blanchet}},
  \bibinfo{author}{\bibfnamefont{A.}~\bibnamefont{Buonanno}}, \bibnamefont{and}
  \bibinfo{author}{\bibfnamefont{G.}~\bibnamefont{Faye}},
  \bibinfo{journal}{Phys.Rev.} \textbf{\bibinfo{volume}{D74}},
  \bibinfo{pages}{104034} (\bibinfo{year}{2006}).

\bibitem[{\citenamefont{Marsat et~al.}(2013)\citenamefont{Marsat, Bohe, Faye,
  and Blanchet}}]{Marsat:2012fn}
\bibinfo{author}{\bibfnamefont{S.}~\bibnamefont{Marsat}},
  \bibinfo{author}{\bibfnamefont{A.}~\bibnamefont{Bohe}},
  \bibinfo{author}{\bibfnamefont{G.}~\bibnamefont{Faye}}, \bibnamefont{and}
  \bibinfo{author}{\bibfnamefont{L.}~\bibnamefont{Blanchet}},
  \bibinfo{journal}{Class.Quantum Grav.} \textbf{\bibinfo{volume}{30}},
  \bibinfo{pages}{055007} (\bibinfo{year}{2013}), \eprint{1210.4143}.

\bibitem[{\citenamefont{Bohe et~al.}(2013{\natexlab{a}})\citenamefont{Bohe,
  Marsat, and Blanchet}}]{Bohe:2013cla}
\bibinfo{author}{\bibfnamefont{A.}~\bibnamefont{Bohe}},
  \bibinfo{author}{\bibfnamefont{S.}~\bibnamefont{Marsat}}, \bibnamefont{and}
  \bibinfo{author}{\bibfnamefont{L.}~\bibnamefont{Blanchet}}
  (\bibinfo{year}{2013}{\natexlab{a}}), \eprint{1303.7412}.

\bibitem[{\citenamefont{Bohe et~al.}(2013{\natexlab{b}})\citenamefont{Bohe,
  Marsat, Faye, and Blanchet}}]{Bohe:2012mr}
\bibinfo{author}{\bibfnamefont{A.}~\bibnamefont{Bohe}},
  \bibinfo{author}{\bibfnamefont{S.}~\bibnamefont{Marsat}},
  \bibinfo{author}{\bibfnamefont{G.}~\bibnamefont{Faye}}, \bibnamefont{and}
  \bibinfo{author}{\bibfnamefont{L.}~\bibnamefont{Blanchet}},
  \bibinfo{journal}{Class.Quant.Grav.} \textbf{\bibinfo{volume}{30}},
  \bibinfo{pages}{075017} (\bibinfo{year}{2013}{\natexlab{b}}),
  \eprint{1212.5520}.

\bibitem[{\citenamefont{Sturani et~al.}(2010)\citenamefont{Sturani, Fischetti,
  Cadonati, Guidi, Healy et~al.}}]{Sturani:2010ju}
\bibinfo{author}{\bibfnamefont{R.}~\bibnamefont{Sturani}},
  \bibinfo{author}{\bibfnamefont{S.}~\bibnamefont{Fischetti}},
  \bibinfo{author}{\bibfnamefont{L.}~\bibnamefont{Cadonati}},
  \bibinfo{author}{\bibfnamefont{G.}~\bibnamefont{Guidi}},
  \bibinfo{author}{\bibfnamefont{J.}~\bibnamefont{Healy}}, \bibnamefont{et~al.}
  (\bibinfo{year}{2010}), \eprint{1012.5172}.

\bibitem[{\citenamefont{Schmidt et~al.}(2012)\citenamefont{Schmidt, Hannam, and
  Husa}}]{Schmidt:2012rh}
\bibinfo{author}{\bibfnamefont{P.}~\bibnamefont{Schmidt}},
  \bibinfo{author}{\bibfnamefont{M.}~\bibnamefont{Hannam}}, \bibnamefont{and}
  \bibinfo{author}{\bibfnamefont{S.}~\bibnamefont{Husa}},
  \bibinfo{journal}{Phys.Rev.} \textbf{\bibinfo{volume}{D86}},
  \bibinfo{pages}{104063} (\bibinfo{year}{2012}), \eprint{1207.3088}.

\bibitem[{\citenamefont{Hannam et~al.}(2013)\citenamefont{Hannam, Schmidt,
  BohŽ, Haegel, Husa et~al.}}]{Hannam:2013oca}
\bibinfo{author}{\bibfnamefont{M.}~\bibnamefont{Hannam}},
  \bibinfo{author}{\bibfnamefont{P.}~\bibnamefont{Schmidt}},
  \bibinfo{author}{\bibfnamefont{A.}~\bibnamefont{BohŽ}},
  \bibinfo{author}{\bibfnamefont{L.}~\bibnamefont{Haegel}},
  \bibinfo{author}{\bibfnamefont{S.}~\bibnamefont{Husa}}, \bibnamefont{et~al.}
  (\bibinfo{year}{2013}), \eprint{1308.3271}.

\bibitem[{\citenamefont{Taracchini et~al.}(2013)\citenamefont{Taracchini,
  Buonanno, Pan, Hinderer, Boyle et~al.}}]{Taracchini:2013rva}
\bibinfo{author}{\bibfnamefont{A.}~\bibnamefont{Taracchini}},
  \bibinfo{author}{\bibfnamefont{A.}~\bibnamefont{Buonanno}},
  \bibinfo{author}{\bibfnamefont{Y.}~\bibnamefont{Pan}},
  \bibinfo{author}{\bibfnamefont{T.}~\bibnamefont{Hinderer}},
  \bibinfo{author}{\bibfnamefont{M.}~\bibnamefont{Boyle}}, \bibnamefont{et~al.}
  (\bibinfo{year}{2013}), \eprint{1311.2544}.

\bibitem[{\citenamefont{Lundgren and O'Shaughnessy}(2013)}]{Lundgren:2013jla}
\bibinfo{author}{\bibfnamefont{A.}~\bibnamefont{Lundgren}} \bibnamefont{and}
  \bibinfo{author}{\bibfnamefont{R.}~\bibnamefont{O'Shaughnessy}}
  (\bibinfo{year}{2013}), \eprint{1304.3332}.

\bibitem[{\citenamefont{Schmidt et~al.}(2011)\citenamefont{Schmidt, Hannam,
  Husa, and Ajith}}]{Schmidt:2010it}
\bibinfo{author}{\bibfnamefont{P.}~\bibnamefont{Schmidt}},
  \bibinfo{author}{\bibfnamefont{M.}~\bibnamefont{Hannam}},
  \bibinfo{author}{\bibfnamefont{S.}~\bibnamefont{Husa}}, \bibnamefont{and}
  \bibinfo{author}{\bibfnamefont{P.}~\bibnamefont{Ajith}},
  \bibinfo{journal}{Phys.Rev.} \textbf{\bibinfo{volume}{D84}},
  \bibinfo{pages}{024046} (\bibinfo{year}{2011}), \eprint{1012.2879}.

\bibitem[{\citenamefont{Boyle et~al.}(2011)\citenamefont{Boyle, Owen, and
  Pfeiffer}}]{Boyle:2011gg}
\bibinfo{author}{\bibfnamefont{M.}~\bibnamefont{Boyle}},
  \bibinfo{author}{\bibfnamefont{R.}~\bibnamefont{Owen}}, \bibnamefont{and}
  \bibinfo{author}{\bibfnamefont{H.~P.} \bibnamefont{Pfeiffer}},
  \bibinfo{journal}{Phys.Rev.} \textbf{\bibinfo{volume}{D84}},
  \bibinfo{pages}{124011} (\bibinfo{year}{2011}), \eprint{1110.2965}.

\bibitem[{\citenamefont{O'Shaughnessy et~al.}(2011)\citenamefont{O'Shaughnessy,
  Vaishnav, Healy, Meeks, and Shoemaker}}]{O'Shaughnessy:2011fx}
\bibinfo{author}{\bibfnamefont{R.}~\bibnamefont{O'Shaughnessy}},
  \bibinfo{author}{\bibfnamefont{B.}~\bibnamefont{Vaishnav}},
  \bibinfo{author}{\bibfnamefont{J.}~\bibnamefont{Healy}},
  \bibinfo{author}{\bibfnamefont{Z.}~\bibnamefont{Meeks}}, \bibnamefont{and}
  \bibinfo{author}{\bibfnamefont{D.}~\bibnamefont{Shoemaker}},
  \bibinfo{journal}{Phys.Rev.} \textbf{\bibinfo{volume}{D84}},
  \bibinfo{pages}{124002} (\bibinfo{year}{2011}), \eprint{1109.5224}.

\bibitem[{\citenamefont{Ochsner and O'Shaughnessy}(2012)}]{Ochsner:2012dj}
\bibinfo{author}{\bibfnamefont{E.}~\bibnamefont{Ochsner}} \bibnamefont{and}
  \bibinfo{author}{\bibfnamefont{R.}~\bibnamefont{O'Shaughnessy}},
  \bibinfo{journal}{Phys.Rev.} \textbf{\bibinfo{volume}{D86}},
  \bibinfo{pages}{104037} (\bibinfo{year}{2012}), \eprint{1205.2287}.

\bibitem[{\citenamefont{Boyle}(2013)}]{Boyle:2013nka}
\bibinfo{author}{\bibfnamefont{M.}~\bibnamefont{Boyle}},
  \bibinfo{journal}{Phys.Rev.} \textbf{\bibinfo{volume}{D87}},
  \bibinfo{pages}{104006} (\bibinfo{year}{2013}), \eprint{1302.2919}.

\bibitem[{\citenamefont{Harry et~al.}(2013)\citenamefont{Harry, Nitz, Brown,
  Lundgren, Ochsner et~al.}}]{Harry:2013tca}
\bibinfo{author}{\bibfnamefont{I.}~\bibnamefont{Harry}},
  \bibinfo{author}{\bibfnamefont{A.}~\bibnamefont{Nitz}},
  \bibinfo{author}{\bibfnamefont{D.~A.} \bibnamefont{Brown}},
  \bibinfo{author}{\bibfnamefont{A.}~\bibnamefont{Lundgren}},
  \bibinfo{author}{\bibfnamefont{E.}~\bibnamefont{Ochsner}},
  \bibnamefont{et~al.} (\bibinfo{year}{2013}), \eprint{1307.3562}.

\bibitem[{\citenamefont{{LIGO Scientific Collaboration}}()}]{LAL}
\bibinfo{author}{\bibnamefont{{LIGO Scientific Collaboration}}},
  \eprint{https://www.lsc-group.phys.uwm.edu/daswg/projects/lal.html}.

\bibitem[{\citenamefont{Aasi et~al.}(2013{\natexlab{a}})}]{Aasi:2013jjl}
\bibinfo{author}{\bibfnamefont{J.}~\bibnamefont{Aasi}} \bibnamefont{et~al.}
  (\bibinfo{collaboration}{LIGO Collaboration, Virgo Collaboration})
  (\bibinfo{year}{2013}{\natexlab{a}}), \eprint{1304.1775}.

\bibitem[{\citenamefont{{van der Sluys} et~al.}(2008)\citenamefont{{van der
  Sluys}, {R{\"o}ver}, {Stroeer}, {Raymond}, {Mandel}, {Christensen},
  {Kalogera}, {Meyer}, and {Vecchio}}}]{2008ApJ...688L..61V}
\bibinfo{author}{\bibfnamefont{M.~V.} \bibnamefont{{van der Sluys}}},
  \bibinfo{author}{\bibfnamefont{C.}~\bibnamefont{{R{\"o}ver}}},
  \bibinfo{author}{\bibfnamefont{A.}~\bibnamefont{{Stroeer}}},
  \bibinfo{author}{\bibfnamefont{V.}~\bibnamefont{{Raymond}}},
  \bibinfo{author}{\bibfnamefont{I.}~\bibnamefont{{Mandel}}},
  \bibinfo{author}{\bibfnamefont{N.}~\bibnamefont{{Christensen}}},
  \bibinfo{author}{\bibfnamefont{V.}~\bibnamefont{{Kalogera}}},
  \bibinfo{author}{\bibfnamefont{R.}~\bibnamefont{{Meyer}}}, \bibnamefont{and}
  \bibinfo{author}{\bibfnamefont{A.}~\bibnamefont{{Vecchio}}},
  \bibinfo{journal}{apjl} \textbf{\bibinfo{volume}{688}}, \bibinfo{pages}{L61}
  (\bibinfo{year}{2008}), \eprint{0710.1897}.

\bibitem[{\citenamefont{{Raymond} et~al.}(2009)\citenamefont{{Raymond}, {van
  der Sluys}, {Mandel}, {Kalogera}, {R{\"o}ver}, and
  {Christensen}}}]{2009CQGra..26k4007R}
\bibinfo{author}{\bibfnamefont{V.}~\bibnamefont{{Raymond}}},
  \bibinfo{author}{\bibfnamefont{M.~V.} \bibnamefont{{van der Sluys}}},
  \bibinfo{author}{\bibfnamefont{I.}~\bibnamefont{{Mandel}}},
  \bibinfo{author}{\bibfnamefont{V.}~\bibnamefont{{Kalogera}}},
  \bibinfo{author}{\bibfnamefont{C.}~\bibnamefont{{R{\"o}ver}}},
  \bibnamefont{and}
  \bibinfo{author}{\bibfnamefont{N.}~\bibnamefont{{Christensen}}},
  \bibinfo{journal}{Classical and Quantum Gravity}
  \textbf{\bibinfo{volume}{26}}, \bibinfo{eid}{114007} (\bibinfo{year}{2009}),
  \eprint{0812.4302}.

\bibitem[{\citenamefont{{O'Shaughnessy}
  et~al.}(2013)\citenamefont{{O'Shaughnessy}, {Farr}, {Ochsner}, {Cho}, {Kim},
  and {Lee}}}]{2013arXiv1308.4704O}
\bibinfo{author}{\bibfnamefont{R.}~\bibnamefont{{O'Shaughnessy}}},
  \bibinfo{author}{\bibfnamefont{B.}~\bibnamefont{{Farr}}},
  \bibinfo{author}{\bibfnamefont{E.}~\bibnamefont{{Ochsner}}},
  \bibinfo{author}{\bibfnamefont{H.-S.} \bibnamefont{{Cho}}},
  \bibinfo{author}{\bibfnamefont{C.}~\bibnamefont{{Kim}}}, \bibnamefont{and}
  \bibinfo{author}{\bibfnamefont{C.-H.} \bibnamefont{{Lee}}},
  \bibinfo{journal}{ArXiv e-prints}  (\bibinfo{year}{2013}),
  \eprint{1308.4704}.

\bibitem[{\citenamefont{{O'Shaughnessy}
  et~al.}(2014)\citenamefont{{O'Shaughnessy}, {Farr}, {Ochsner}, {Cho},
  {Raymond}, {Kim}, and {Lee}}}]{2014arXiv1403.0544O}
\bibinfo{author}{\bibfnamefont{R.}~\bibnamefont{{O'Shaughnessy}}},
  \bibinfo{author}{\bibfnamefont{B.}~\bibnamefont{{Farr}}},
  \bibinfo{author}{\bibfnamefont{E.}~\bibnamefont{{Ochsner}}},
  \bibinfo{author}{\bibfnamefont{H.~S.} \bibnamefont{{Cho}}},
  \bibinfo{author}{\bibfnamefont{V.}~\bibnamefont{{Raymond}}},
  \bibinfo{author}{\bibfnamefont{C.}~\bibnamefont{{Kim}}}, \bibnamefont{and}
  \bibinfo{author}{\bibfnamefont{C.~H.} \bibnamefont{{Lee}}},
  \bibinfo{journal}{ArXiv e-prints}  (\bibinfo{year}{2014}),
  \eprint{1403.0544}.

\bibitem[{\citenamefont{{Vitale} et~al.}(2014)\citenamefont{{Vitale}, {Lynch},
  {Veitch}, {Raymond}, and {Sturani}}}]{2014arXiv1403.0129V}
\bibinfo{author}{\bibfnamefont{S.}~\bibnamefont{{Vitale}}},
  \bibinfo{author}{\bibfnamefont{R.}~\bibnamefont{{Lynch}}},
  \bibinfo{author}{\bibfnamefont{J.}~\bibnamefont{{Veitch}}},
  \bibinfo{author}{\bibfnamefont{V.}~\bibnamefont{{Raymond}}},
  \bibnamefont{and}
  \bibinfo{author}{\bibfnamefont{R.}~\bibnamefont{{Sturani}}},
  \bibinfo{journal}{ArXiv e-prints}  (\bibinfo{year}{2014}),
  \eprint{1403.0129}.

\bibitem[{\citenamefont{Buonanno et~al.}(2009)\citenamefont{Buonanno, Iyer,
  Ochsner, Pan, and Sathyaprakash}}]{Buonanno:2009zt}
\bibinfo{author}{\bibfnamefont{A.}~\bibnamefont{Buonanno}},
  \bibinfo{author}{\bibfnamefont{B.~R.} \bibnamefont{Iyer}},
  \bibinfo{author}{\bibfnamefont{E.}~\bibnamefont{Ochsner}},
  \bibinfo{author}{\bibfnamefont{Y.}~\bibnamefont{Pan}}, \bibnamefont{and}
  \bibinfo{author}{\bibfnamefont{B.~S.} \bibnamefont{Sathyaprakash}},
  \bibinfo{journal}{Phys. Rev.} \textbf{\bibinfo{volume}{D80}},
  \bibinfo{pages}{084043} (\bibinfo{year}{2009}), \eprint{0907.0700}.

\bibitem[{\citenamefont{Pan et~al.}(2004)\citenamefont{Pan, Buonanno, Chen, and
  Vallisneri}}]{Pan:2003qt}
\bibinfo{author}{\bibfnamefont{Y.}~\bibnamefont{Pan}},
  \bibinfo{author}{\bibfnamefont{A.}~\bibnamefont{Buonanno}},
  \bibinfo{author}{\bibfnamefont{Y.-b.} \bibnamefont{Chen}}, \bibnamefont{and}
  \bibinfo{author}{\bibfnamefont{M.}~\bibnamefont{Vallisneri}},
  \bibinfo{journal}{Phys.Rev.} \textbf{\bibinfo{volume}{D69}},
  \bibinfo{pages}{104017} (\bibinfo{year}{2004}), \eprint{gr-qc/0310034}.

\bibitem[{\citenamefont{Buonanno et~al.}(2004)\citenamefont{Buonanno, Chen,
  Pan, and Vallisneri}}]{Buonanno:2004yd}
\bibinfo{author}{\bibfnamefont{A.}~\bibnamefont{Buonanno}},
  \bibinfo{author}{\bibfnamefont{Y.-b.} \bibnamefont{Chen}},
  \bibinfo{author}{\bibfnamefont{Y.}~\bibnamefont{Pan}}, \bibnamefont{and}
  \bibinfo{author}{\bibfnamefont{M.}~\bibnamefont{Vallisneri}},
  \bibinfo{journal}{Phys.Rev.} \textbf{\bibinfo{volume}{D70}},
  \bibinfo{pages}{104003} (\bibinfo{year}{2004}), \eprint{gr-qc/0405090}.

\bibitem[{\citenamefont{Nitz et~al.}(2013)\citenamefont{Nitz, Lundgren, Brown,
  Ochsner, Keppel et~al.}}]{Nitz:2013mxa}
\bibinfo{author}{\bibfnamefont{A.~H.} \bibnamefont{Nitz}},
  \bibinfo{author}{\bibfnamefont{A.}~\bibnamefont{Lundgren}},
  \bibinfo{author}{\bibfnamefont{D.~A.} \bibnamefont{Brown}},
  \bibinfo{author}{\bibfnamefont{E.}~\bibnamefont{Ochsner}},
  \bibinfo{author}{\bibfnamefont{D.}~\bibnamefont{Keppel}},
  \bibnamefont{et~al.} (\bibinfo{year}{2013}), \eprint{1307.1757}.

\bibitem[{\citenamefont{O'Shaughnessy et~al.}(2013)\citenamefont{O'Shaughnessy,
  Farr, Ochsner, Cho, Kim et~al.}}]{O'Shaughnessy:2013vma}
\bibinfo{author}{\bibfnamefont{R.}~\bibnamefont{O'Shaughnessy}},
  \bibinfo{author}{\bibfnamefont{B.}~\bibnamefont{Farr}},
  \bibinfo{author}{\bibfnamefont{E.}~\bibnamefont{Ochsner}},
  \bibinfo{author}{\bibfnamefont{H.-S.} \bibnamefont{Cho}},
  \bibinfo{author}{\bibfnamefont{C.}~\bibnamefont{Kim}}, \bibnamefont{et~al.}
  (\bibinfo{year}{2013}), \eprint{1308.4704}.

\bibitem[{\citenamefont{O'Shaughnessy et~al.}(2014)\citenamefont{O'Shaughnessy,
  Farr, Ochsner, Cho, Raymond et~al.}}]{O'Shaughnessy:2014dka}
\bibinfo{author}{\bibfnamefont{R.}~\bibnamefont{O'Shaughnessy}},
  \bibinfo{author}{\bibfnamefont{B.}~\bibnamefont{Farr}},
  \bibinfo{author}{\bibfnamefont{E.}~\bibnamefont{Ochsner}},
  \bibinfo{author}{\bibfnamefont{H.}~\bibnamefont{Cho}},
  \bibinfo{author}{\bibfnamefont{V.}~\bibnamefont{Raymond}},
  \bibnamefont{et~al.} (\bibinfo{year}{2014}), \eprint{1403.0544}.

\bibitem[{\citenamefont{Schnittman}(2004)}]{Schnittman:2004vq}
\bibinfo{author}{\bibfnamefont{J.~D.} \bibnamefont{Schnittman}},
  \bibinfo{journal}{Phys.Rev.} \textbf{\bibinfo{volume}{D70}},
  \bibinfo{pages}{124020} (\bibinfo{year}{2004}), \eprint{astro-ph/0409174}.

\bibitem[{\citenamefont{{Metropolis} et~al.}(1953)\citenamefont{{Metropolis},
  {Rosenbluth}, {Rosenbluth}, {Teller}, and {Teller}}}]{metropolis53}
\bibinfo{author}{\bibfnamefont{N.}~\bibnamefont{{Metropolis}}},
  \bibinfo{author}{\bibfnamefont{A.~W.} \bibnamefont{{Rosenbluth}}},
  \bibinfo{author}{\bibfnamefont{M.~N.} \bibnamefont{{Rosenbluth}}},
  \bibinfo{author}{\bibfnamefont{A.~H.} \bibnamefont{{Teller}}},
  \bibnamefont{and} \bibinfo{author}{\bibfnamefont{E.}~\bibnamefont{{Teller}}},
  \bibinfo{journal}{\jcp} \textbf{\bibinfo{volume}{21}}, \bibinfo{pages}{1087}
  (\bibinfo{year}{1953}).

\bibitem[{\citenamefont{Hastings}(1970)}]{hastings}
\bibinfo{author}{\bibfnamefont{W.~K.} \bibnamefont{Hastings}},
  \bibinfo{journal}{Biometrika} \textbf{\bibinfo{volume}{57}},
  \bibinfo{pages}{97} (\bibinfo{year}{1970}).

\bibitem[{\citenamefont{Gelman et~al.}(1996)\citenamefont{Gelman, Roberts, and
  Gilks}}]{Gelman96}
\bibinfo{author}{\bibfnamefont{A.}~\bibnamefont{Gelman}},
  \bibinfo{author}{\bibfnamefont{G.~O.} \bibnamefont{Roberts}},
  \bibnamefont{and} \bibinfo{author}{\bibfnamefont{W.~R.} \bibnamefont{Gilks}},
  in \emph{\bibinfo{booktitle}{Bayesian statistics, 5 (Alicante, 1994)}}
  (\bibinfo{publisher}{Oxford Univ. Press}, \bibinfo{address}{New York},
  \bibinfo{year}{1996}), Oxford Sci. Publ., pp. \bibinfo{pages}{599--607}.

\bibitem[{\citenamefont{Aasi et~al.}(2013{\natexlab{b}})}]{Aasi:2013wya}
\bibinfo{author}{\bibfnamefont{J.}~\bibnamefont{Aasi}} \bibnamefont{et~al.}
  (\bibinfo{collaboration}{LIGO Scientific Collaboration, Virgo Collaboration})
  (\bibinfo{year}{2013}{\natexlab{b}}), \eprint{1304.0670}.

\bibitem[{\citenamefont{{Manchester} et~al.}(2005)\citenamefont{{Manchester},
  {Hobbs}, {Teoh}, and {Hobbs}}}]{2005yCat.7245....0M}
\bibinfo{author}{\bibfnamefont{R.~N.} \bibnamefont{{Manchester}}},
  \bibinfo{author}{\bibfnamefont{G.~B.} \bibnamefont{{Hobbs}}},
  \bibinfo{author}{\bibfnamefont{A.}~\bibnamefont{{Teoh}}}, \bibnamefont{and}
  \bibinfo{author}{\bibfnamefont{M.}~\bibnamefont{{Hobbs}}},
  \bibinfo{journal}{VizieR Online Data Catalog}
  \textbf{\bibinfo{volume}{7245}}, \bibinfo{pages}{0} (\bibinfo{year}{2005}).

\bibitem[{\citenamefont{{Cannon} et~al.}(2012)\citenamefont{{Cannon}, {Cariou},
  {Chapman}, {Crispin-Ortuzar}, {Fotopoulos}, {Frei}, {Hanna}, {Kara},
  {Keppel}, {Liao} et~al.}}]{2012ApJ...748..136C}
\bibinfo{author}{\bibfnamefont{K.}~\bibnamefont{{Cannon}}},
  \bibinfo{author}{\bibfnamefont{R.}~\bibnamefont{{Cariou}}},
  \bibinfo{author}{\bibfnamefont{A.}~\bibnamefont{{Chapman}}},
  \bibinfo{author}{\bibfnamefont{M.}~\bibnamefont{{Crispin-Ortuzar}}},
  \bibinfo{author}{\bibfnamefont{N.}~\bibnamefont{{Fotopoulos}}},
  \bibinfo{author}{\bibfnamefont{M.}~\bibnamefont{{Frei}}},
  \bibinfo{author}{\bibfnamefont{C.}~\bibnamefont{{Hanna}}},
  \bibinfo{author}{\bibfnamefont{E.}~\bibnamefont{{Kara}}},
  \bibinfo{author}{\bibfnamefont{D.}~\bibnamefont{{Keppel}}},
  \bibinfo{author}{\bibfnamefont{L.}~\bibnamefont{{Liao}}},
  \bibnamefont{et~al.}, \bibinfo{journal}{\apj} \textbf{\bibinfo{volume}{748}},
  \bibinfo{eid}{136} (\bibinfo{year}{2012}), \eprint{1107.2665}.

\bibitem[{\citenamefont{{Singer} et~al.}(2014)}]{first2years}
\bibinfo{author}{\bibfnamefont{L.}~\bibnamefont{{Singer}}}
  \bibnamefont{et~al.}, \bibinfo{journal}{in prep}  (\bibinfo{year}{2014}).

\bibitem[{\citenamefont{Baird et~al.}(2013)\citenamefont{Baird, Fairhurst,
  Hannam, and Murphy}}]{Baird:2012cu}
\bibinfo{author}{\bibfnamefont{E.}~\bibnamefont{Baird}},
  \bibinfo{author}{\bibfnamefont{S.}~\bibnamefont{Fairhurst}},
  \bibinfo{author}{\bibfnamefont{M.}~\bibnamefont{Hannam}}, \bibnamefont{and}
  \bibinfo{author}{\bibfnamefont{P.}~\bibnamefont{Murphy}},
  \bibinfo{journal}{Phys.Rev.} \textbf{\bibinfo{volume}{D87}},
  \bibinfo{pages}{024035} (\bibinfo{year}{2013}), \eprint{1211.0546}.

\bibitem[{\citenamefont{{Hannam} et~al.}(2013)\citenamefont{{Hannam}, {Brown},
  {Fairhurst}, {Fryer}, and {Harry}}}]{2013ApJ...766L..14H}
\bibinfo{author}{\bibfnamefont{M.}~\bibnamefont{{Hannam}}},
  \bibinfo{author}{\bibfnamefont{D.~A.} \bibnamefont{{Brown}}},
  \bibinfo{author}{\bibfnamefont{S.}~\bibnamefont{{Fairhurst}}},
  \bibinfo{author}{\bibfnamefont{C.~L.} \bibnamefont{{Fryer}}},
  \bibnamefont{and} \bibinfo{author}{\bibfnamefont{I.~W.}
  \bibnamefont{{Harry}}}, \bibinfo{journal}{apjl}
  \textbf{\bibinfo{volume}{766}}, \bibinfo{eid}{L14} (\bibinfo{year}{2013}),
  \eprint{1301.5616}.

\bibitem[{\citenamefont{{Aasi} et~al.}(2014)\citenamefont{{Aasi}, {Abbott},
  {Abbott}, {Abbott}, {Abernathy}, {Accadia}, and
  et~al.}}]{2014arXiv1401.0939T}
\bibinfo{author}{\bibfnamefont{J.}~\bibnamefont{{Aasi}}},
  \bibinfo{author}{\bibfnamefont{B.~P.} \bibnamefont{{Abbott}}},
  \bibinfo{author}{\bibfnamefont{R.}~\bibnamefont{{Abbott}}},
  \bibinfo{author}{\bibfnamefont{T.}~\bibnamefont{{Abbott}}},
  \bibinfo{author}{\bibfnamefont{M.~R.} \bibnamefont{{Abernathy}}},
  \bibinfo{author}{\bibfnamefont{T.}~\bibnamefont{{Accadia}}},
  \bibnamefont{and} \bibinfo{author}{\bibnamefont{et~al.}},
  \bibinfo{journal}{ArXiv e-prints}  (\bibinfo{year}{2014}),
  \eprint{1401.0939}.

\bibitem[{\citenamefont{{Chatziioannou}
  et~al.}(2014)\citenamefont{{Chatziioannou}, {Cornish}, {Klein}, and
  {Yunes}}}]{2014arXiv1404.3180C}
\bibinfo{author}{\bibfnamefont{K.}~\bibnamefont{{Chatziioannou}}},
  \bibinfo{author}{\bibfnamefont{N.}~\bibnamefont{{Cornish}}},
  \bibinfo{author}{\bibfnamefont{A.}~\bibnamefont{{Klein}}}, \bibnamefont{and}
  \bibinfo{author}{\bibfnamefont{N.}~\bibnamefont{{Yunes}}},
  \bibinfo{journal}{ArXiv e-prints}  (\bibinfo{year}{2014}),
  \eprint{1404.3180}.

\bibitem[{\citenamefont{{Pan} et~al.}(2014)\citenamefont{{Pan}, {Buonanno},
  {Taracchini}, {Kidder}, {Mrou{\'e}}, {Pfeiffer}, {Scheel}, and
  {Szil{\'a}gyi}}}]{2014PhRvD..89h4006P}
\bibinfo{author}{\bibfnamefont{Y.}~\bibnamefont{{Pan}}},
  \bibinfo{author}{\bibfnamefont{A.}~\bibnamefont{{Buonanno}}},
  \bibinfo{author}{\bibfnamefont{A.}~\bibnamefont{{Taracchini}}},
  \bibinfo{author}{\bibfnamefont{L.~E.} \bibnamefont{{Kidder}}},
  \bibinfo{author}{\bibfnamefont{A.~H.} \bibnamefont{{Mrou{\'e}}}},
  \bibinfo{author}{\bibfnamefont{H.~P.} \bibnamefont{{Pfeiffer}}},
  \bibinfo{author}{\bibfnamefont{M.~A.} \bibnamefont{{Scheel}}},
  \bibnamefont{and}
  \bibinfo{author}{\bibfnamefont{B.}~\bibnamefont{{Szil{\'a}gyi}}},
  \bibinfo{journal}{\prd} \textbf{\bibinfo{volume}{89}}, \bibinfo{eid}{084006}
  (\bibinfo{year}{2014}), \eprint{1307.6232}.

\end{thebibliography}
\end{document}